\documentclass{article}[12pt]
\usepackage{amsmath,amssymb,siunitx,graphicx,subcaption,color,cite}
\usepackage[a4paper]{geometry}
\usepackage{hyperref}

\newcommand{\cp}{\mathbf{cp}}
\newcommand{\e}{\mathbf{e}}
\newcommand{\mbfn}{\mathbf{n}}
\newcommand{\mbfr}{\mathbf{r}}
\newcommand{\mbfx}{\mathbf{x}}
\newcommand{\mbfu}{\mathbf{u}}

\newcommand{\mbfrho}{\boldsymbol{\rho}}

\newcommand{\mbft}{\boldsymbol{\tau}}

\title{The Curvature-Augmented Closest Point Method with Vesicle Inextensibility Application}
\date{}
\author{Christopher J. Vogl \footnotemark[2]}

\begin{document}
\maketitle

\renewcommand{\thefootnote}{\fnsymbol{footnote}}

\footnotetext[2]{Center for Applied Scientific Computing, Lawrence Livermore National Laboratory, Livermore, CA 94550 (chris.j.vogl@gmail.com).}

\begin{abstract}
The Closest Point method, initially developed by Ruuth and Merriman, allows for the numerical solution of surface partial differential equations without the need for a parameterization of the surface itself.  Surface quantities are embedded into the surrounding domain by assigning each value at a given spatial location to the corresponding value at the closest point on the surface.  This embedding allows for surface derivatives to be replaced by their Cartesian counterparts (e.g. $\nabla_s = \nabla$).  This equivalence is only valid on the surface, and thus, interpolation is used to enforce what is known as the side condition away from the surface.  To improve upon the method, this work derives an operator embedding that incorporates curvature information, making it valid in a neighborhood of the surface.  With this, direct enforcement of the side condition is no longer needed.  Comparisons in $\mathbb{R}^2$ and $\mathbb{R}^3$ show that the resulting Curvature-Augmented Closest Point method has better accuracy and requires less memory, through increased matrix sparsity, than the Closest Point method, while maintaining similar matrix condition numbers.  To demonstrate the utility of the method in a physical application, simulations of inextensible, bi-lipid vesicles evolving toward equilibrium shapes are also included.
\end{abstract}

%%%%%%%%%%%%%%%%%%%%%
%%%%%%%%%%%%%%%%%%%%%
\section{Introduction}

Surface partial differential equations (PDEs) are prevalent throughout the scientific literature, including diffusion across a surface \cite{wen_generalized_2014, eloul_charge_2015} and tension along a surface \cite{salac_reynolds_2012,vogl_effect_2014}.  The Closest Point method was developed by Ruuth and Merriman \cite{ruuth_simple_2008} as a numerical method for solving such PDEs by embedding them into the surrounding Cartesian space.  This allows for numerical approximation and solution using standard finite-difference schemes while avoiding the need for an explicit parameterization of the surface.  Since its original development, the Closest Point method has been improved in various ways, including the use of WENO interpolation by Macdonald and Ruuth \cite{macdonald_level_2008}, use of implicit time integrators by Macdonald and Ruuth \cite{macdonald_implicit_2009}, and application of multigrid solvers for elliptic PDEs by Chen and Macdonald \cite{chen_closest_2015}.
\par
This work continues to improve the Closest Point method by generalizing the embedding of surface operators in the PDE.  Currently, surface operators are embedded using expressions that are only valid on the surface.  Interpolation operators are then included at all points to complete the embedding off the surface.  By including curvature information, these expressions are replaced by ones that are valid in a neighborhood of the surface.  The result is that interpolation is only needed as a computational boundary condition for the embedding domain.  The improved method, named the Curvature-Augmented Closest Point method, shows increased accuracy of the resulting numerical solution, at least for the test cases within, and increased sparsity of the resulting linear system.
\par
A summary of the Closest Point method is first provided, including the operator embedding expressions for the surface gradient and Laplace-Beltrami operators.  The generalized embedding expressions for the Curvature-Augmented Closest Point method are then derived for one-dimensional surfaces in $\mathbb{R}^2$ and for two-dimensional surfaces in $\mathbb{R}^3$.  For the latter, specific expressions for a sphere and for axisymmetric surfaces are found.  The Closest Point and Curvature-Augmented Closest Point methods are then compared to analytic solutions for simple test cases in $\mathbb{R}^2$ and $\mathbb{R}^3$.  Finally, the Curvature-Augmented Closest Point method is applied to inextensible vesicles as they relax to different equilibrium shapes.

%%%%%%%%%%%%%%%%%%%%%
%%%%%%%%%%%%%%%%%%%%%
\section{Formulation}

As this work seeks to improve the Closest Point (CP) method, it is motivated in a similar fashion as in Chen and Macdonald \cite{chen_closest_2015}.  Consider a smooth, closed surface $\Gamma \subset \mathbb{R}^{n+1}$ of dimension $n$ defined by $\mbfr(\sigma_1,\ldots,\sigma_n)$.  On this surface, define the elliptic PDE
\begin{align}
	\tilde{c}(\mbfx)\tilde{\gamma}(\mbfx) - \Delta_s\tilde{\gamma}(\mbfx) = \tilde{m}(\mbfx), \quad \mbfx \in \Gamma, \label{EQ:elliptic_SPDE}
\end{align}
governing some surface quantity of interest $\tilde{\gamma}(\mbfx)$.  Here, $\tilde{c}(\mbfx)$ and $\tilde{m}(\mbfx)$ are known surface fields.  Given that $\Gamma$ is an $n$-dimensional Riemann manifold, the Laplace-Beltrami operator $\Delta_s$ has the general form
\begin{align}
	\Delta_s f = \frac{1}{\sqrt{|g|}} (\sqrt{|g|} g^{ij} f_{\sigma_j})_{\sigma_i},
\end{align}
where $g_{ij} := \mbfr_{\sigma_i} \cdot \mbfr_{\sigma_j}$ is the metric tensor, $|g|$ is the determinant of $g$, and $g^{ij}$ are the components of $g^{-1}$.  If $n=1$, then $\Delta_s \tilde{\gamma} = \tilde{\gamma}_{ss}$, where $\partial / \partial s := |\mbfr_\sigma|^{-1} \partial / \partial \sigma$.  This can be numerically approximated using a relatively standard centered-difference formula for the variable-coefficient Laplace operator given a discrete parameterization: $\mbfr(\sigma_i)$.  If $n>1$, however, the form of $\Delta_s \tilde{\gamma}$ can become significantly more complicated depending on the structure of $\mbfr(\sigma_1,\ldots,\sigma_n)$.  Furthermore, constructing a discrete parameterization for an arbitrary surface that avoids singularities can prove to be difficult or intractable for $n > 1$.  As an example, for $n=2$, consider that $\mbfr_\theta \rightarrow \mathbf{0}$ as $\psi \rightarrow \pm \pi/2$ for a sphere defined by $\mbfr(\theta,\psi) = \cos(\theta)\cos(\psi) \mathbf{i} + \sin(\theta)\cos(\psi) \mathbf{j} + \cos(\psi) \mathbf{k}$, where $\mathbf{i}$, $\mathbf{j}$, $\mathbf{k}$ form the standard Cartesian basis for $\mathbf{R}^3$.
\par
If the CP method is used instead to solve for $\tilde{\gamma}$, the surface fields are first embedded in $\mathbb{R}^{n+1}$ via $c(\mbfx) := \tilde{c}(\cp(\mbfx))$ and $m(\mbfx) := \tilde{m}(\cp(\mbfx))$, where $\cp$: $\mathbb{R}^{n+1} \rightarrow \Gamma$ such that $\cp(\mbfx)$ is the closest point on $\Gamma$ to $\mbfx$.  An unknown field $\gamma(\mbfx)$, which will eventually contain the solution for $\tilde{\gamma}$, is introduced with the property that $\gamma(\mbfx) = \gamma(\cp(\mbfx))$.  The CP method  uses the key relations that
\begin{align}
	\nabla_s \gamma(\mbfx) = \nabla \gamma(\mbfx) \quad \text{ and } \quad \Delta_s \gamma(\mbfx) = \Delta \gamma(\mbfx), \quad \mbfx \in \Gamma, \label{EQ:LB_relation}
\end{align}
where $\nabla_s \gamma := (I - \mbfn \otimes \mbfn) \nabla \gamma$, $\mbfn$ is the outward-pointing normal to $\Gamma$, and $\otimes$ is the tensor outer product.  Thus, $\gamma$ is found by embedding $\Delta_s \gamma$ as $\Delta \gamma$ and solving the alternative PDE
\begin{align*}
	c(\mbfx)\gamma(\mbfx) - \Delta \gamma(\mbfx)= m(\mbfx), \quad \mbfx \in \mathbb{R}^n, \text{ (consistency condition)}\\
	\text{subject to } \gamma(\mbfx) = \gamma(\cp(\mbfx)). \text{ (side condition)}
\end{align*}
Note the above PDE is invariant to how $\Gamma$ is parameterized.  As for solving it, there are various approaches to combining finite-difference and interpolation schemes \cite{ruuth_simple_2008,macdonald_implicit_2009}.  The more recent approach by Chen and Macdonald \cite{chen_closest_2015} involves using a discrete Laplacian matrix operator $L$ and $q^\text{th}$ order interpolation matrix operators $E_q$:
\begin{align*}
	A \boldsymbol{\gamma} = \mathbf{m}, \text{ where } A = I - M \text{ and } M = E_{q_1} L - \omega (I - E_{q_2}).
\end{align*}
As Chen and Macdonald note, the parameter $\omega$ balances the consistency and side conditions.  Finally, because the solution $\gamma(\mbfx)$ also satisfies (\ref{EQ:elliptic_SPDE}), the final solution is obtained as $\tilde{\gamma}(\mbfx) = \gamma(\mbfx)$ for $\mbfx \in \Gamma$.
\par
Now that the CP method is introduced, the focus of this work is improving the accuracy of the numerical solution and sparsity of the resulting linear system in solving the surface PDE (\ref{EQ:elliptic_SPDE}), without losing either the invariance to surface parameterization or the ability to use existing Laplacian and interpolation stencils.  This is accomplished by including curvature information that generalizes (\ref{EQ:LB_relation}) to a neighborhood about the surface, resulting in better accuracy.  Interpolation is then only needed to close the embedded domain, resulting in a more sparse linear system.  This improved approach is named the Curvature-Augmented Closest Point (CACP) method.

%%%%%%%%%%%%%%%%%%%%%
\subsection{CACP method in $\mathbb{R}^2$}

For $n=1$, let $\Gamma$ be defined by $\mbfr(\sigma)$.  Construct a coordinate system for a neighborhood $\Omega_\Gamma$ about $\Gamma$ via $\mbfx(\sigma,\eta) = \mbfr(\sigma) + \eta \mbfn(\sigma)$.  Denote the local curvature as $\kappa(\sigma) = \mbfn_\sigma(\sigma) \cdot \mbfr_\sigma(\sigma) / |\mbfr_\sigma(\sigma)|^2$.  This all allows for a compact expression of the metric tensor and its inverse in this coordinate system:
\begin{align*}
	g = \left[ \begin{array}{cc} (1 + \eta \kappa)^2|\mbfr_\sigma|^2 & 0 \\ 0 & 1 \end{array} \right] \text{ and } g^{-1} =  \left[ \begin{array}{cc} \frac{1}{(1 + \eta \kappa)^2|\mbfr_\sigma|^2} & 0 \\ 0 & 1 \end{array} \right].
\end{align*}
Restrict $\eta$ so that $1 + \eta \kappa \ne 0$ for all of $\Omega_\Gamma$.  This gives the following definitions for the gradient and divergence operators in this coordinate system:
\begin{align*}
	\nabla f &:= g^{ij} f_{\sigma_j} \mbfx_{\sigma_i} = \frac{1}{(1 + \eta \kappa) |\mbfr_\sigma|}f_\sigma \mbft + f_\eta \mbfn,\\
	\nabla \cdot \mathbf{f} &:= \frac{1}{\sqrt{|g|}}\left( \frac{ \sqrt{|g|}}{\sqrt{g_{ii}}} \mathbf{f} \cdot \frac{\mbfr_{\sigma_i}}{\sqrt{g_{ii}}} \right )_{\sigma_i} = \frac{1}{(1 + \eta \kappa)}\left[ \frac{1}{|\mbfr_\sigma|}(\mathbf{f} \cdot \mbft)_\sigma + \big((1 + \eta \kappa) \mathbf{f} \cdot \mbfn \big)_\eta \right],
\end{align*}
where $\mbft$ is the tangent vector $\mbfr_\sigma/|\mbfr_\sigma|$.
\par
Denote $\phi(\mbfx)$ as the signed-distance level set function, where $\phi(\mbfx(\sigma,\eta)) = \eta$, and a global curvature $\kappa(\mbfx)$ such that $\kappa(\mbfx(\sigma,\eta)) = \kappa(\sigma)$.  Recalling that $\gamma(\mbfx) = \gamma(\cp(\mbfx))$, so that $\gamma_\eta = 0$, a generalization of the surface gradient and  Laplace-Beltrami embeddings (\ref{EQ:LB_relation}) are obtained for $\Omega_\Gamma$:
\begin{align}
	\begin{aligned}
		&(1 + \phi(\mbfx) \kappa(\mbfx)) \nabla \gamma = \frac{1}{|\mbfr_\sigma|} \gamma_\sigma \mbft := \nabla_s \gamma \quad \text{and} \\
		&(1 + \phi(\mbfx) \kappa(\mbfx)) \nabla \cdot \big[ \big( (1 + \phi(\mbfx)\kappa(\mbfx)\big) \nabla \gamma(\mbfx) \big] = \frac{1}{|\mbfr_\sigma|}\left(\frac{1}{|\mbfr_\sigma|} \gamma_\sigma \right)_\sigma := \Delta _s \gamma.
	\end{aligned}
\label{EQ:LB_relation2D}
\end{align}
Note that (\ref{EQ:LB_relation2D}) reduces to (\ref{EQ:LB_relation}) if $\mbfx$ is restricted to $\Gamma$.
\par
A different alternative PDE is now used for $\gamma(\mbfx)$:
\begin{align*}
	(1 + \phi(\mbfx)\kappa(\mbfx)) \nabla \cdot ((1 + \phi(\mbfx)\kappa(\mbfx))\nabla \gamma(\mbfx)) - c(\mbfx)\gamma(\mbfx) = f(\mbfx), \quad \mbfx \in \Omega_\Gamma.
\end{align*}
As with the CP method, this alternative PDE is invariant to how $\Gamma$ is parameterized.  Additionally, the extra structure in this PDE eliminates the need to explicitly satisfy the side condition (see Appendix \ref{APP:side_condition}).  Thus, the solution to $\gamma$, and consequently $\tilde{\gamma}$, is found using standard finite-difference formulas for the variable-coefficient Laplacian, with interpolation stencils only needed to close the corresponding system $A \boldsymbol{\gamma} = \mathbf{m}$.

%%%%%%%%%%%%%%%%%%%%%
\subsection{CACP method in $\mathbb{R}^3$}

For $n=2$, consider a $\Gamma$ where local parameterizations exist along the lines of principal curvature.  In other words, given a point $\mathbf{y}$ on $\Gamma$, a parameterization $\mbfr(\sigma_1,\sigma_2)$ exists such that $\mbfr_{\sigma_1}$ and $\mbfr_{\sigma_2}$ are the principal directions for all points in a neighborhood of $\mathbf{y}$ on $\Gamma$.  Note that this means $\mbfr_{\sigma_1} \cdot \mbfr_{\sigma_2} = 0$, giving the following for the surface gradient and Laplace-Beltrami operators:
\begin{align*}
	&\nabla_s f = \frac{1}{|\mbfr_{\sigma_1}|} f_{\sigma_1} \mbft^{\sigma_1} + \frac{1}{|\mbfr_{\sigma_2}|} f_{\sigma_2}\mbft^{\sigma_2} \quad \text{and} \\
	&\Delta_s f = \frac{1}{|\mbfr_{\sigma_1}||\mbfr_{\sigma_2}|}\left[ \left(\frac{|\mbfr_{\sigma_2}|}{|\mbfr_{\sigma_1}|} f_{\sigma_1}\right)_{\sigma_1} + \left(\frac{|\mbfr_{\sigma_2}|}{|\mbfr_{\sigma_1}|} f_{\sigma_2}\right)_{\sigma_2}\right],
\end{align*}
where $\mbft^{\sigma_i} = \mbfr_{\sigma_i}/|\mbfr_{\sigma_i}|$.  Again, the goal is to generalize (\ref{EQ:LB_relation}) for the operators above.  Similar to the previous section, for a neighborhood $\Omega_\Gamma$ about $\Gamma$, introduce the following coordinate system for $\mathbb{R}^3$: $\mbfx(\sigma_1,\sigma_2,\eta) = \mbfr(\sigma_1,\sigma_2) + \eta \mbfn(\sigma_1,\sigma_2)$.  Denote the local curvatures $\kappa(\sigma_1,\sigma_2) = \mbfn_{\sigma_1} \cdot \mbfr_{\sigma_1} / |\mbfr_{\sigma_1}|^2$ and $h(\sigma_1,\sigma_2) = \mbfn_{\sigma_2} \cdot \mbfr_{\sigma_2} / |\mbfr_{\sigma_2}|^2$, which are also the principal curvatures.  Note that the principal directions also diagonalize the shape tensor ($\mbfn_{\sigma_j} \cdot \mbfr_{\sigma_i} = -\mbfr_{\sigma_i \sigma_j} \cdot \mbfn = 0$ for $i \ne j$).  This gives the compact form for the metric tensor, in this coordinate system, as
\begin{align*}
	g = \left[ \begin{array}{ccc} [(1 + \eta \kappa)|\mbfr_{\sigma_1}|]^2 & 0 & 0 \\ 0 & [(1 + \eta h)|\mbfr_{\sigma_2}|]^2 & 0 \\ 0 & 0 & 1 \end{array} \right ],
\end{align*}
with the form for $g^{-1}$ following easily.
\par
Let $T$ be a second-order tensor such that $T(\sigma_1,\sigma_2,\eta) = a \, \mbft^{\sigma_1} \otimes \mbft^{\sigma_1} + b \, \mbft^{\sigma_2} \otimes \mbft^{\sigma_2}$.  Assuming a function $f$ is such that $f(\mbfx) = f(\cp(\mbfx))$, the following result is true for gradient and divergence operations:
\begin{align*}
	\nabla f =& \frac{1}{(1 + \eta \kappa) |\mbfr_{\sigma_1}|} f_{\sigma_1} \mbft^{\sigma_1}  + \frac{1}{(1 + \eta h) |\mbfr_{\sigma_2}|} f_{\sigma_2} \mbft^{\sigma_2} \\
	\Rightarrow \nabla \cdot (T \nabla f ) =& \frac{1}{(1 + \eta \kappa)(1+ \eta h)|\mbfr_{\sigma_1}||\mbfr_{\sigma_2}|} \\
&\left[ \big( \frac{1 + \eta h}{1 + \eta \kappa}a \frac{|\mbfr_{\sigma_2}|}{|\mbfr_{\sigma_1}|}f_{\sigma_1} \big)_{\sigma_1} + \big(\frac{1 + \eta \kappa}{1 + \eta h} b\frac{|\mbfr_{\sigma_1}|}{|\mbfr_{\sigma_2}|} f_{\sigma_2}  \big)_{\sigma_2} \right],
\end{align*}
Again denote $\phi(\mbfx)$ as the signed-distance level set function, $\kappa(\mbfx)$ as one global curvature so that $\kappa(\mbfx(\sigma_1,\sigma_2,\eta)) = \kappa(\sigma_1,\sigma_2)$, $h(\mbfx)$ as the other global curvature so that $h(\mbfx(\sigma_1,\sigma_2,\eta)) = h(\sigma_1,\sigma_2)$, and $\mbft^{\sigma_i}(\mbfx)$ such that $\mbft^{\sigma_i}(\mbfx(\sigma_1,\sigma_2,\eta)) = \mbft^{\sigma_i}(\sigma_1,\sigma_2)$.  Additionally, define  the following two tensors:
\begin{align*}
	S(\mbfx(\sigma_1,\sigma_2,\eta)) &= (1 + \phi(\mbfx)\kappa(\mbfx))  \big(\mbft^{\sigma_1} \otimes \mbft^{\sigma_1}\big)(\mbfx)+ (1 + \phi(\mbfx)h(\mbfx)) \big( \mbft^{\sigma_2} \otimes \mbft^{\sigma_2} \big)(\mbfx), \\
	T(\mbfx(\sigma_1,\sigma_2,\eta)) &= \frac{1 + \phi(\mbfx)\kappa(\mbfx)}{1 + \phi(\mbfx)h(\mbfx)} \big(\mbft^{\sigma_1} \otimes \mbft^{\sigma_1}\big)(\mbfx)+ \frac{1 + \phi(\mbfx)h(\mbfx)}{1 + \phi(\mbfx)\kappa(\mbfx)} \big( \mbft^{\sigma_2} \otimes \mbft^{\sigma_2} \big)(\mbfx).
\end{align*}
A generalization of (\ref{EQ:LB_relation}) is now available for $\gamma(\mbfx) = \gamma(\cp(\mbfx))$:
\begin{align}
	\nabla_s \gamma = S(\mbfx) \nabla \gamma \quad \text{and} \quad \Delta_s \gamma = (1 + \phi(\mbfx)\kappa(\mbfx))(1 + \phi(\mbfx)h(\mbfx)) \nabla \cdot \big( T(\mbfx) \nabla \gamma \big), \label{EQ:LB_relation3D}
\end{align}
that latter of which is used in an alternative PDE for $\gamma$, similar to $n=1$. Again, the alternative PDE is invariant to how $\Gamma$ is parameterized, because the generalized embedding (\ref{EQ:LB_relation3D}) only depends on the principal directions and curvatures, which are invariant to parameterization.

\subsubsection{Example: a sphere in $\mathbf{R}^3$}

Consider $\Gamma$ to be a sphere of radius $R$ that is defined by $\mbfr(\theta,\psi) = R \mbfrho(\theta,\psi)$, where $\mbfrho(\theta,\psi) = \cos(\theta)\cos(\psi)\mathbf{i} + \sin(\theta)\cos(\psi)\mathbf{j} + \sin(\psi)\mathbf{k}$, $\theta \in [0,2\pi)$, and $\psi \in [-\pi, \pi]$.  Note that $\mbfr_\theta$ and $\mbfr_\psi$ are principal directions for all points on $\Gamma$ except at $\psi = \pm \pi/2$.  The corresponding principal curvatures are both $1/R$.  Thus, the gradient and Laplace-Beltrami operator embeddings (\ref{EQ:LB_relation3D}) are
\begin{align}
	\nabla_s \gamma = (1 + \phi(\mbfx)/R) \nabla \gamma \quad \text{and} \quad \Delta_s \gamma = (1 + \phi(\mbfx)/R)^2 \nabla \cdot \big( \nabla \gamma \big), \label{EQ:LB_relation3D_sphere}
\end{align}
because $S(\mbfx) = (1 + \phi(\mbfx)/R)I$ and $T(\mbfx) = I$.  Again, note that (\ref{EQ:LB_relation3D_sphere}) is unaffected by how $\mbfr_\theta \rightarrow \mathbf{0}$ as $\psi \rightarrow \pm \pi/2$.

\subsubsection{Example: axisymmetric surface in $\mathbf{R}^3$} \label{SEC:improvement_R3_axis}

Consider $\Gamma$ to be an axisymmetric shape that is defined by $\mbfr(\sigma,\theta) = x(\sigma)\e_x(\theta) + y(\sigma)\e_y$, where $\e_x(\theta) = \cos(\theta)\mathbf{i} + \sin(\theta)\mathbf{j}$, $\e_y = \mathbf{k}$, and $\theta \in [0,2\pi)$.  Define $\e_\theta(\theta) = \e_x'(\theta)$ and note the following:
\begin{align*}
	\mbfn &:= \frac{\mbfr_\theta \times \mbfr_\sigma}{|\mbfr_\theta \times \mbfr_\sigma|} = \frac{y_\sigma \e_x - x_\sigma \e_y}{\sqrt{x_\sigma^2 + y_\sigma^2}},  \quad\quad \mbfr_{\sigma \theta} \cdot \mbfn = 0 \\
	\kappa &= -\frac{\mbfr_{\sigma \sigma} \cdot \mbfn}{|\mbfr_\sigma|^2} = -\frac{x_{\sigma\sigma}y_\sigma - y_{\sigma\sigma}x_\sigma}{(x_\sigma^2 + y_\sigma^2)^{3/2}}, \quad\quad h = -\frac{\mbfr_{\theta\theta} \cdot \mbfn}{|\mbfr_\theta|^2} = \frac{y_\sigma}{x \sqrt{x_\sigma^2 + y_\sigma^2}}.
\end{align*}
Thus, $\mbfr_\sigma$ and $\mbfr_\theta$ are principal directions with $\kappa$ and $h$ as the corresponding principal curvatures.  Thus, the surface gradient and Laplace-Beltrami operator embeddings (\ref{EQ:LB_relation3D}) can be applied with $\mbft^{\sigma} = (x_\sigma \e_x(\theta) + y_\sigma \e_y)/\sqrt{x_\sigma^2 + y_\sigma^2}$ and $\mbft^{\theta} = \e_\theta(\theta)$.  If $\tilde{c}(\mbfx)$ and $\tilde{m}(\mbfx)$ are also axisymmetric, as is the case with vesicle surface tension explored later, the embeddings are more compact because one assumes that $\gamma(\sigma,\theta) = \gamma(\sigma)$ :
\begin{align}
	\begin{aligned}
		\nabla_s \gamma &= (1 + \phi(\mbfx)\kappa(\mbfx)) \nabla \gamma \quad \text{and} \\
		\Delta_s \gamma &= (1 + \phi(\mbfx)\kappa(\mbfx))(1 + \phi(\mbfx)h(\mbfx)) \nabla \cdot \left ( \frac{1 + \phi(\mbfx)\kappa(\mbfx)}{1 + \phi(\mbfx)h(\mbfx)} \nabla \gamma\right).
	\end{aligned}
	\label{EQ:LB_relation3D_axis}
\end{align}

%%%%%%%%%%%%%%%%%%%%%
%%%%%%%%%%%%%%%%%%%%%
\section{Comparing the CP and CACP Methods}

In order to improve upon the Closest Point (CP) method in solving an elliptic PDE (\ref{EQ:elliptic_SPDE}) on a surface, curvature-augmented embeddings of the surface gradient and Laplace-Beltrami operators are derived for one-dimensional surfaces in $\mathbb{R}^2$ (\ref{EQ:LB_relation2D}) and for two-dimensional surfaces in $\mathbb{R}^3$ (\ref{EQ:LB_relation3D}), given the availability of principal directions and curvatures.  A specific version of (\ref{EQ:LB_relation3D}) is derived for a sphere (\ref{EQ:LB_relation3D_sphere}).  To compare this Curvature-Augmented Closest Point (CACP) method with the original CP method, PDEs are solved on a circle in $\mathbb{R}^2$, a clover surface in $\mathbb{R}^2$, and a sphere in $\mathbb{R}^3$, where analytic solutions are known for each surface.  The accuracy of solution and sparsity of matrices are compared in each case.  The Matlab code used for these comparisons is available at \href{https://github.com/cjvogl/cacp}{https://github.com/cjvogl/cacp}.

%%%%%%%%%%%%%%%%%%%%%
\subsection{Results in $\mathbb{R}^2$}

Following the lead of Chen and Macdonald \cite{chen_closest_2015}, the elliptic PDE (\ref{EQ:elliptic_SPDE}) is chosen so that the analytic solution is $\gamma(\theta) = \sin(\theta) + \sin(12\theta)$ for a given surface $\Gamma$.  The PDE is then solved on the domain $[$-$2,2] \times [$-$2,2]$, which is discretized into $M+1$ grid nodes in each direction: $\mbfx_{i,j} = ($-$2 + i \Delta x,\, $-$2 + j \Delta x)$, $\Delta x = 4/M$.  The grid nodes are then enumerated ($\mbfx_k$) so the discretization can be written as a linear system.
\par
Any grid node that is a part of the bi-cubic interpolation stencil for a point on $\Gamma$ is marked as an \textit{interpolation} node.  Any neighboring node of an \textit{interpolation} node that is not also a \textit{interpolation} node is marked as an \textit{edge} node.  From \cite{chen_closest_2015}, the CP method is implemented as
\begin{align*}
	A \boldsymbol{\gamma} = \mathbf{m}, \text{ where } A = I - M \text{ and } M = E_1 L - \frac{4}{\Delta x^2}(I - E_3).
\end{align*}
Recall that $E_1$ and $E_3$ are bi-linear and bi-cubic interpolation matrices, respectively, and $\mathbf{m}_k = m(\mbfx_k)$.  If $\mbfx_k$ is an \textit{interpolation} node, $L_{k,j}$ corresponds to the standard centered-difference formula for the Laplacian.  If $\mbfx_k$ is an \textit{edge} node, $L_{k,j}$ is arbitrarily set to $\delta_{k,j}$ because $(E_1)_{i,k} = 0$ for all $i$.  For the CACP method, a different form for $A$ is used.  If $\mbfx_k$ is an \textit{edge} node, $A_{k,j} = (4/\Delta x^2)(I - E_3)_{k,j}$.  Note the scaling factor $4/\Delta x^2$ is carried over from the CP method to control the condition number of $A$.  If $\mbfx_k$ is an \textit{interpolation} node, then $A_{k,j}$ corresponds to the variable-coefficient, centered-difference formula applied to (\ref{EQ:LB_relation2D}):
\begin{align*}
	(\Delta_s \gamma)_{i,j} := &\Delta_s \gamma(\mbfx_{i,j}) \approx (1 + \phi_{i,j}\kappa_{i,j}) \\
&\sum_{m \in \{-1,1\}}\left( \alpha_{i+m/2,j}\frac{\gamma_{i+m,j} - \gamma_{i,j}}{\Delta x^2} + \beta_{i,j+m/2}\frac{\gamma_{i,j+m} - \gamma_{i,j}}{\Delta x^2}\right),\\
&\alpha_{i,j} = \sum_{m\in\{-1,1\}}\frac{1 + \phi_{i+m/2,j}\kappa_{i+m/2,j}}{2},\\
&\beta_{i,j} = \sum_{m\in\{-1,1\}}\frac{1 + \phi_{i,j+m/2}\kappa_{i,j+m/2}}{2}.
\end{align*}

\subsubsection{Choosing $\Gamma$ as the unit circle}

With the implementations of the CP and CACP methods defined, the first test surface is the unit circle.  In order to obtain the analytic solution for $\gamma(\theta)$ mentioned above, the PDE is chosen with $\tilde{c}(\theta) = 1$ and $\tilde{m}(\theta) = 2\sin(\theta) + 145\sin(12\theta)$.  Additionally, the signed-distance level set function and curvature for the unit circle are $\phi(\mbfx) = |\mbfx| - 1$ and $\kappa(\mbfx) = 1$.  The second-order accuracy of the CACP method solution, denoted $\gamma_\text{num}(\mbfx)$, is verified against the analytic solution, denoted $\gamma_\text{exact}(\mbfx)$, through both the discrete $L_2$ and $L_\infty$ errors in Table \ref{TAB:CACP_errors_circle}.  Note these errors are defined as
\begin{align*}
	|\boldsymbol{\gamma}_\text{exact} - \boldsymbol{\gamma}_\text{num}|_2 &:= \frac{1}{\tilde{M}}\left[\sum_{k=1}^{\tilde{M}} (\gamma_\text{exact}(\cp(\mbfx_k)) - \gamma_\text{num}(\mbfx_k))^2\right]^{1/2},\\
	|\boldsymbol{\gamma}_\text{exact} - \boldsymbol{\gamma}_\text{num}|_\infty &:=\max_{k=1,\ldots,\tilde{M}} |\gamma_\text{exact}(\cp(\mbfx_k)) - \gamma_\text{num}(\mbfx_k)|,
\end{align*}
where $\cp(\mbfx) = \mbfx - \phi(\mbfx)\nabla\phi(\mbfx)$ and $\tilde{M}$ is the total number of \textit{interpolation} and \textit{edge} nodes.  With the accuracy of the CACP method verified, it is now compared against the CP method for the unit circle.
\begin{table}[h]
	\center
	\begin{tabular}{|l||
		S[round-mode=places,round-precision=4]|
		S[round-mode=places,round-precision=2]|
		S[round-mode=places,round-precision=4]|
		S[round-mode=places,round-precision=2]|}
		\hline
		& {$|\boldsymbol{\gamma}_\text{exact} - \boldsymbol{\gamma}_\text{num}|_2$} & {ratio} & {$|\boldsymbol{\gamma}_\text{exact} - \boldsymbol{\gamma}_\text{num}|_\infty$} & {ratio} \\
		\hline
		{$M = 40$} & 6.529582e-02 & {---} & 1.530827e-01 & {---} \\
		{$M = 80$} & 1.536154e-02 & 4.250604e+00 & 2.676026e-02 & 5.720524e+00 \\
		{$M = 160$} & 3.756485e-03 & 4.089339e+00 &  6.485326e-03 & 4.126278e+00 \\
		{$M = 320$}  & 9.423799e-04 & 3.986168e+00 & 1.593578e-03 & 4.069663e+00 \\
		{$M = 640$}  & 2.358238e-04 & 3.996119e+00 & 3.977083e-04 & 4.006902e+00 \\
		\hline
	\end{tabular}
	\caption{error values for the CACP method on the unit circle in $\mathbb{R}^2$}
	\label{TAB:CACP_errors_circle}
\end{table}
\par
\begin{figure}[h]
	\center
	\begin{subfigure}[t]{0.49\textwidth}
		\center
		\includegraphics[width=\textwidth]{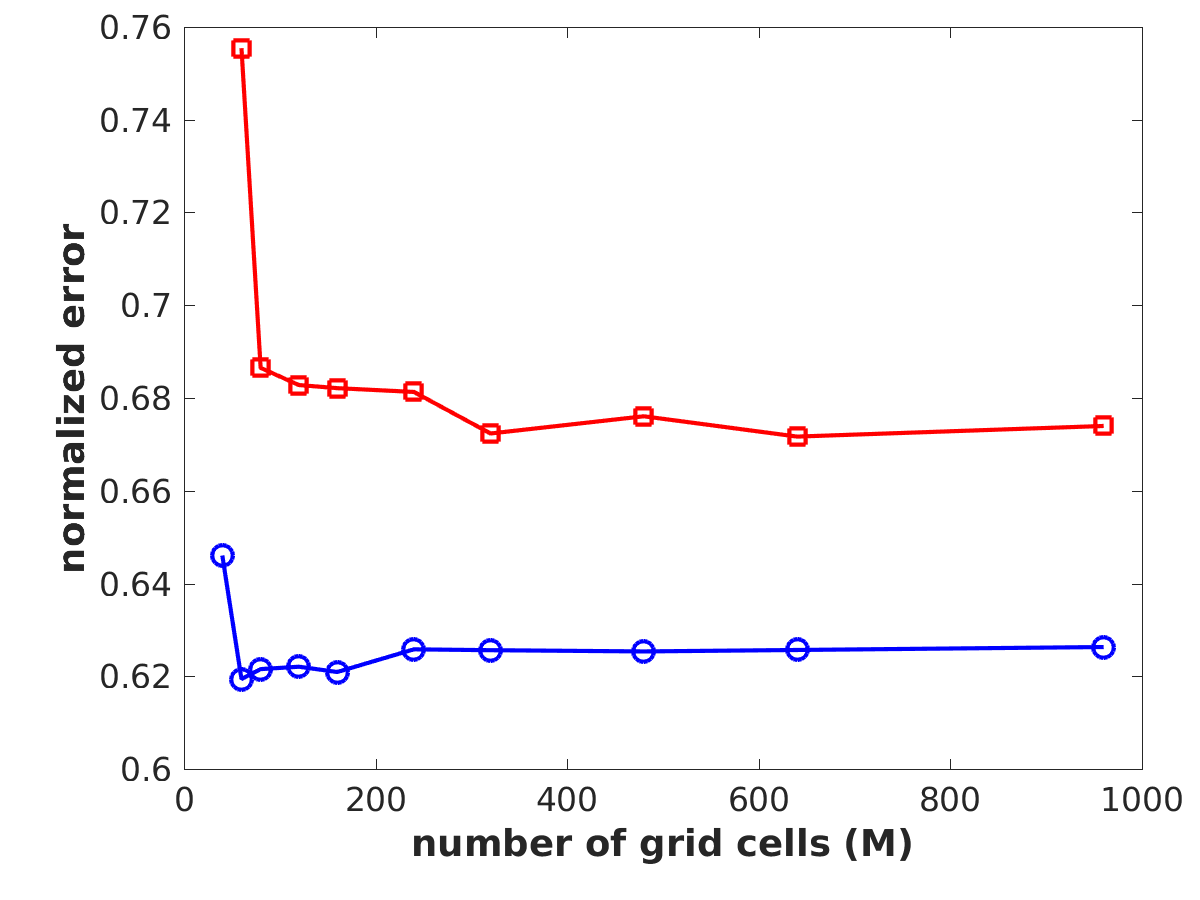}
		\caption{CACP error normalized by CP error (\textcolor{blue}{circles} - $L_2$ error, \textcolor{red}{squares} - $L_\infty$ error)}
		\label{FIG:CP_vs_CACP_circle_errors}
	\end{subfigure}
	\begin{subfigure}[t]{0.49\textwidth}
		\center
		\includegraphics[width=\textwidth]{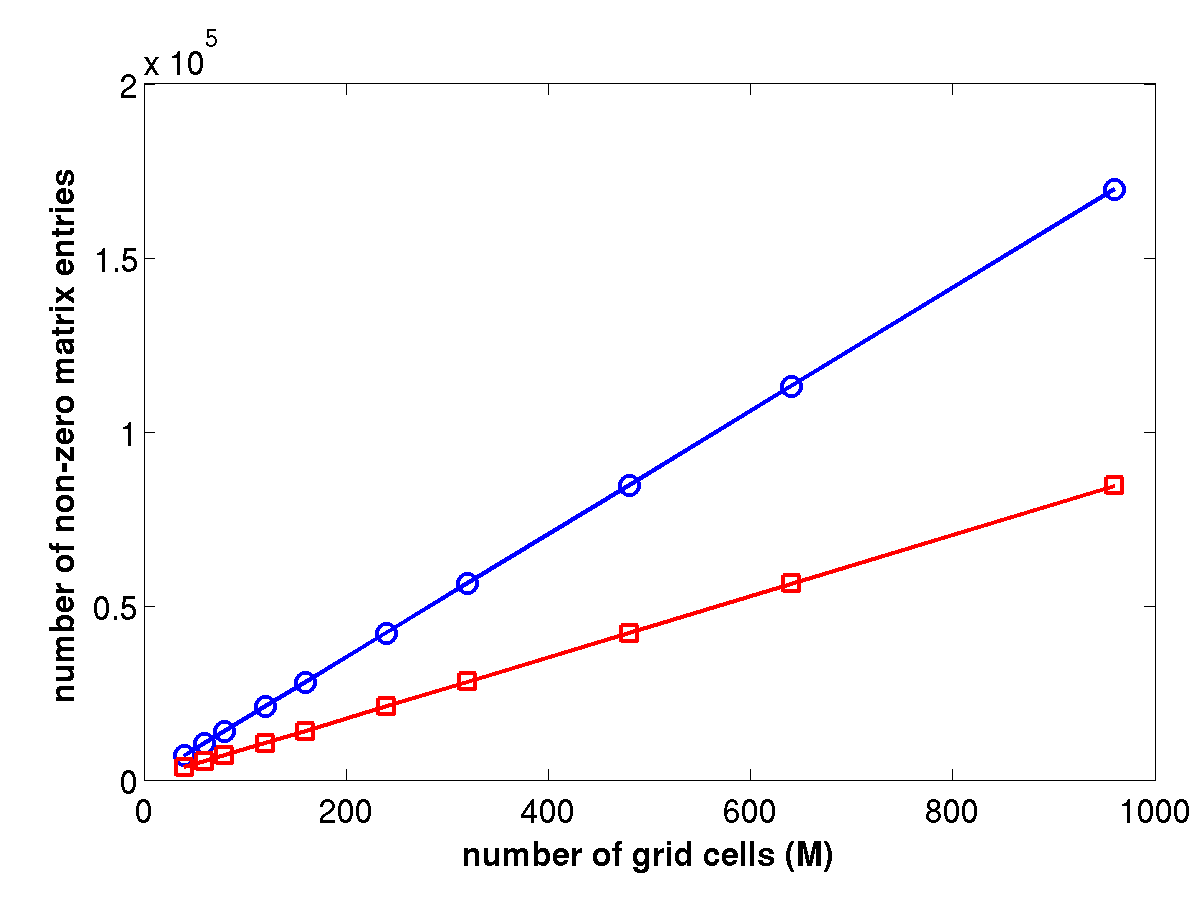}
		\caption{non-zero entries in $A$ (\textcolor{blue}{circles} - CP data with linear fit, \textcolor{red}{squares} - CACP data with linear fit)}
		\label{FIG:CP_vs_CACP_circle_nnz}
	\end{subfigure}
	\caption{comparing CACP method to CP method for the unit circle in $\mathbb{R}^2$}
	\label{FIG:CP_vs_CACP_circle}
\end{figure}

Although the formal accuracy of both the CP and CACP solutions is of second-order, comparing the errors shows the improved accuracy of the CACP method.  Figure \ref{FIG:CP_vs_CACP_circle_errors} shows the error of the CACP method normalized by the error of the CP method.  As more grid nodes are added, the CACP method has an $L_2$ error of about $0.62$ of the CP method and an $L_\infty$ error of about $0.68$ of the CP method.
Additionally, the number of non-zero entries in $A$ is shown for each method in Figure \ref{FIG:CP_vs_CACP_circle_nnz}.  The linear fit in the figure for the CP method results has a slope of approximately $177$, while the slope of the fit for the CACP method results has a slope of approximately $88$.  Thus, when compared to the CP method, the CACP method produces about around a third less error as $M$ increases, while the number of non-zero entries of $A$, and the corresponding memory requirement, grows at about half the rate.  It is important to verify that this increase in sparsity does not come at the cost of matrix conditioning.  Therefore, the matrix condition numbers for both methods, estimated by the Matlab \texttt{condest} routine, are shown in Table \ref{TAB:condition_numbers_circle} of Appendix \ref{APP:condition_numbers} to be the same order of magnitude.

\subsubsection{Choosing $\Gamma$ as the clover}

To ensure that the results in Figure \ref{FIG:CP_vs_CACP_circle} do not depend on $\Gamma$ being a circle, a clover-like shape is now investigated.  The surface is defined by $\mbfr(\theta) = g(\theta)[\cos(\theta)\mathbf{i} + \sin(\theta)\mathbf{j}]$, where $g(\theta) = 1 + 0.25\cos(4\theta -\pi)$.  Thus, in order to illicit a solution for $\gamma(\theta)$ equal to $u(\theta) := \sin(\theta) + \sin(12 \theta)$, the PDE (\ref{EQ:elliptic_SPDE}) is chosen with
\begin{align*}
	 \tilde{c}(\theta) = 1 \quad \text{ and } \quad \tilde{m}(\theta) = -\frac{u_{\theta\theta}}{g_\theta^2 + g^2} + \frac{(g_\theta g_{\theta\theta} + g g_\theta) u_\theta}{(g_\theta^2 + g^2)^2} + u.
\end{align*}
The signed-distance level set and curvature are not readily available in analytic expressions for the clover.  Thus, the closest point on the surface to each grid point is found by minimizing $|\mbfx_{i,j} - \mbfr(\theta)|^2$ over $\theta$, denoting $\theta_{i,j}$ as the minimizer.  Then $\phi(\mbfx_{i,j}) = \pm |\mbfx_{i,j} - \mbfr(\theta_{i,j})|$, with positive values for nodes outside the clover, and $\kappa(\mbfx_{i,j}) = [(g^2 - g_{\theta \theta} g + 2 g_\theta^2)/(g_\theta^2 + g^2)^{3/2}](\mbfx_{i,j})$.
\par
\begin{figure}[h]
	\center
	\begin{subfigure}[t]{0.49\textwidth}
		\center
		\includegraphics[width=\textwidth]{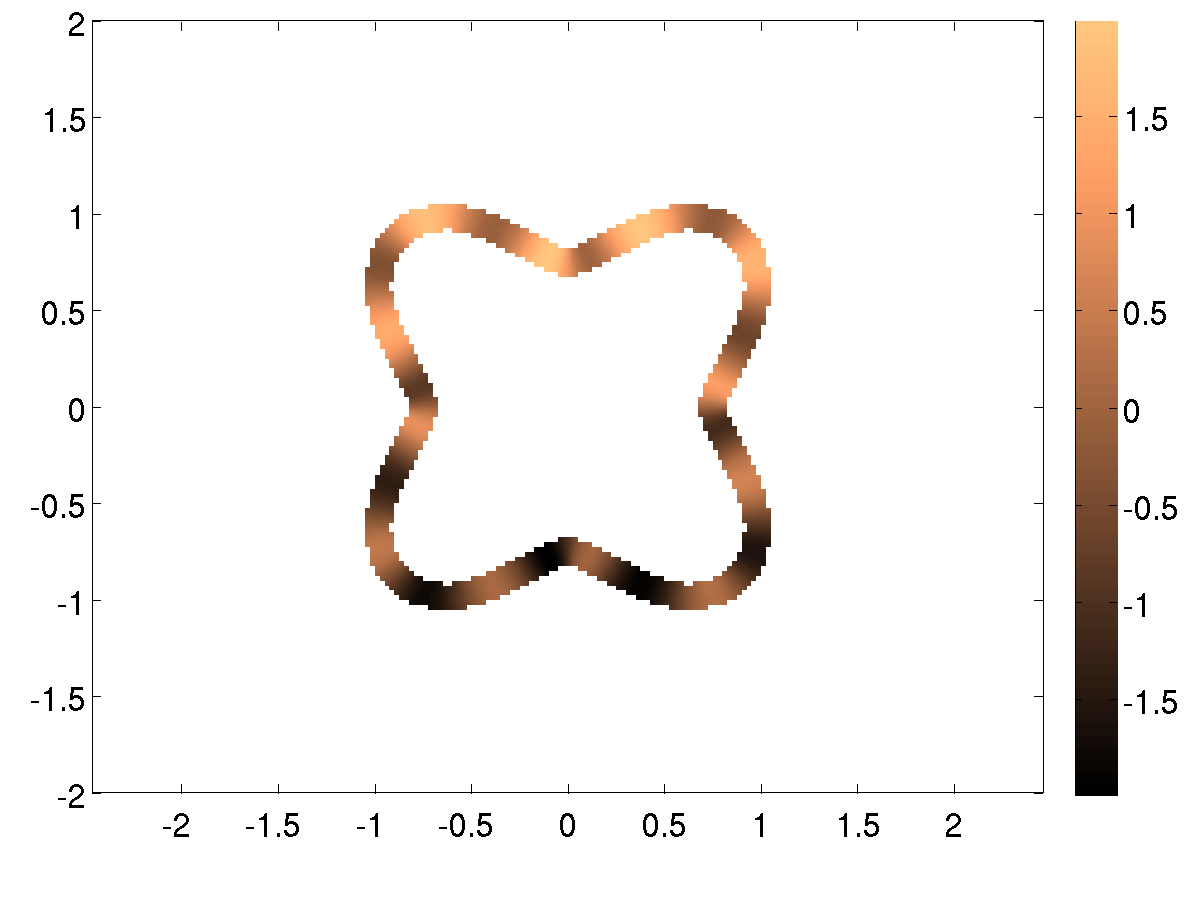}
		\caption{clover shape, with solution, for $M=160$}
		\label{FIG:clover_solution}
	\end{subfigure}
	\begin{subfigure}[t]{0.49\textwidth}
		\center
		\includegraphics[width=\textwidth]{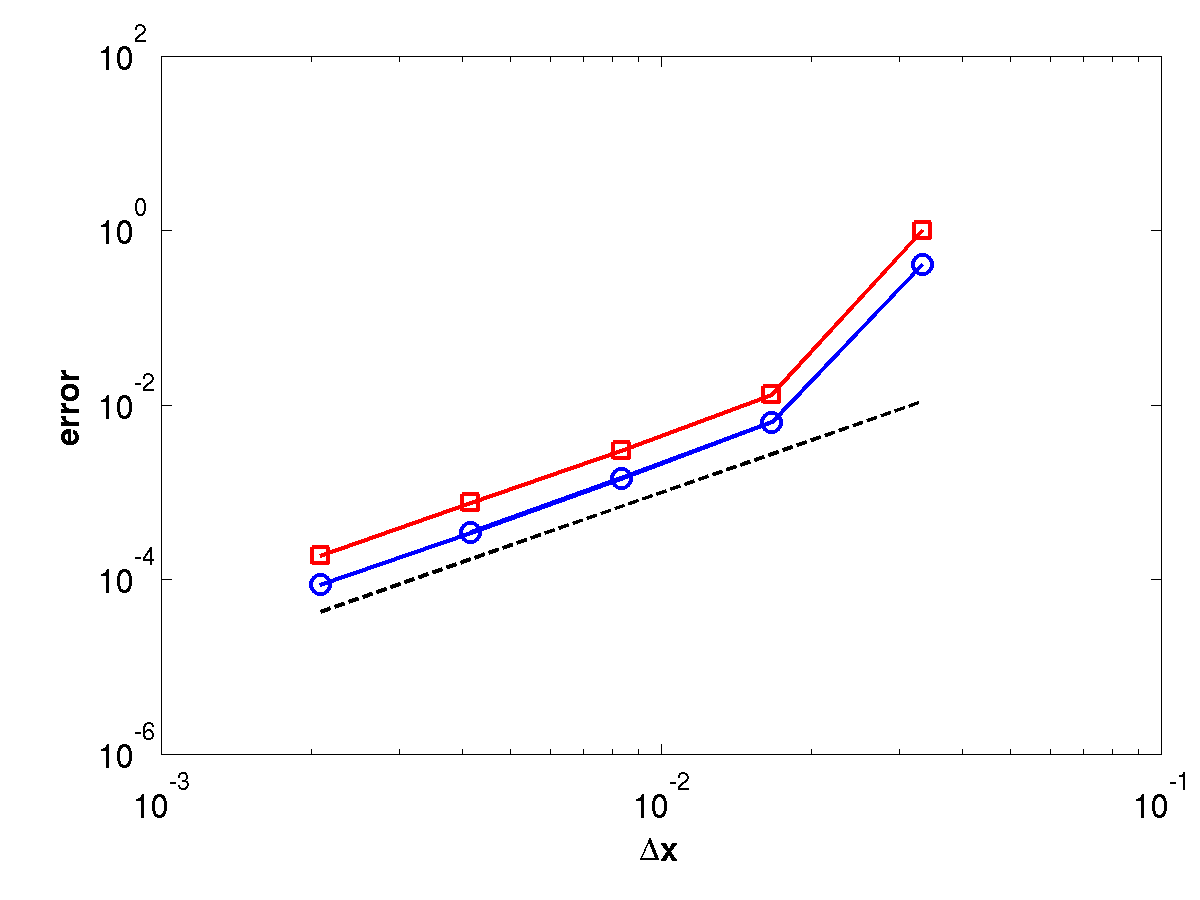}
		\caption{convergence results (\textcolor{blue}{circles} - $L_2$ error, \textcolor{red}{squares} - $L_\infty$ error, dashed - reference $2^\text{nd}$ order line)}
		\label{FIG:clover_convergence}
	\end{subfigure}
	\caption{CACP results for the clover in $\mathbb{R}^2$}
\end{figure}
\begin{figure}[h]
	\center
	\begin{subfigure}[t]{0.49\textwidth}
		\center
		\includegraphics[width=\textwidth]{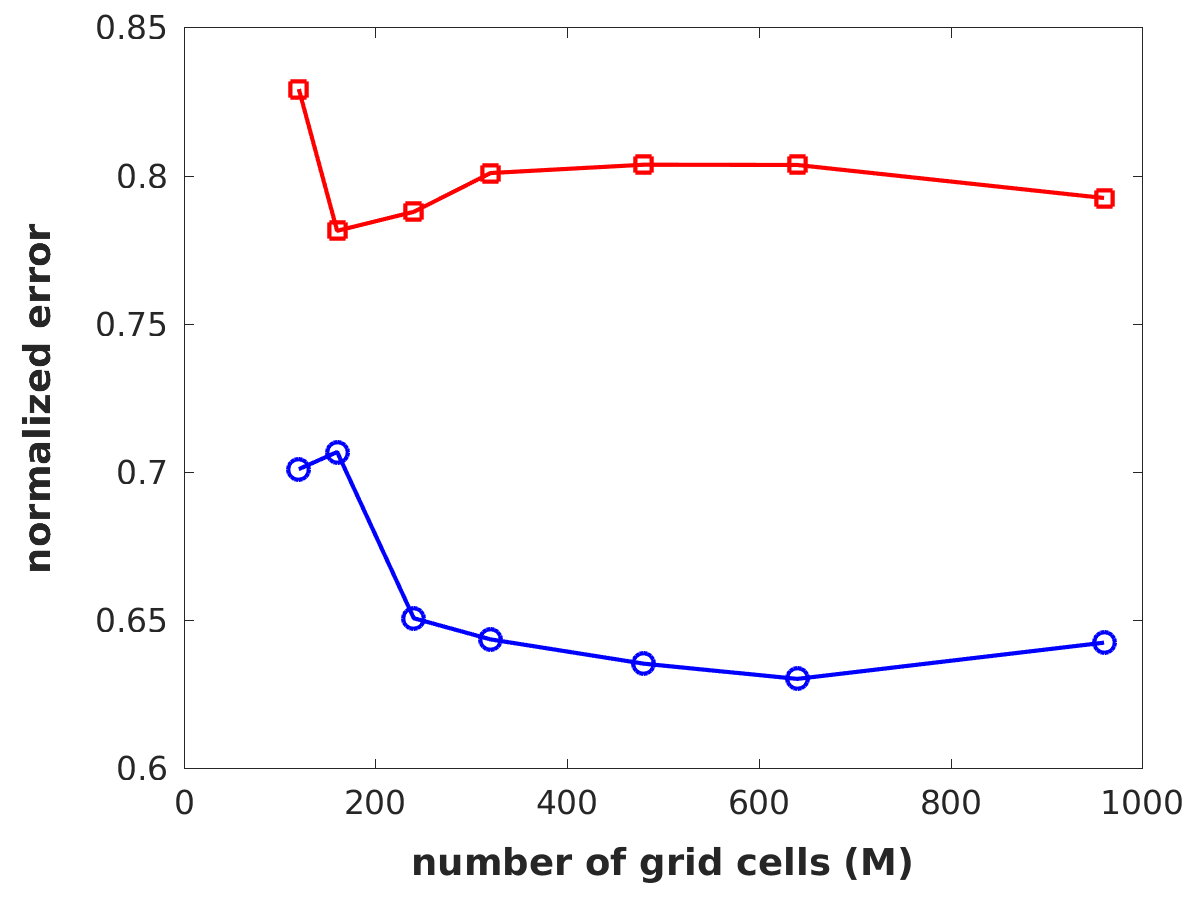}
		\caption{CACP error normalized by CP error (\textcolor{blue}{circles} - $L_2$ error, \textcolor{red}{squares} - $L_\infty$ error)}
		\label{FIG:CP_vs_CACP_clover_errors}
	\end{subfigure}
	\begin{subfigure}[t]{0.49\textwidth}
		\center
		\includegraphics[width=\textwidth]{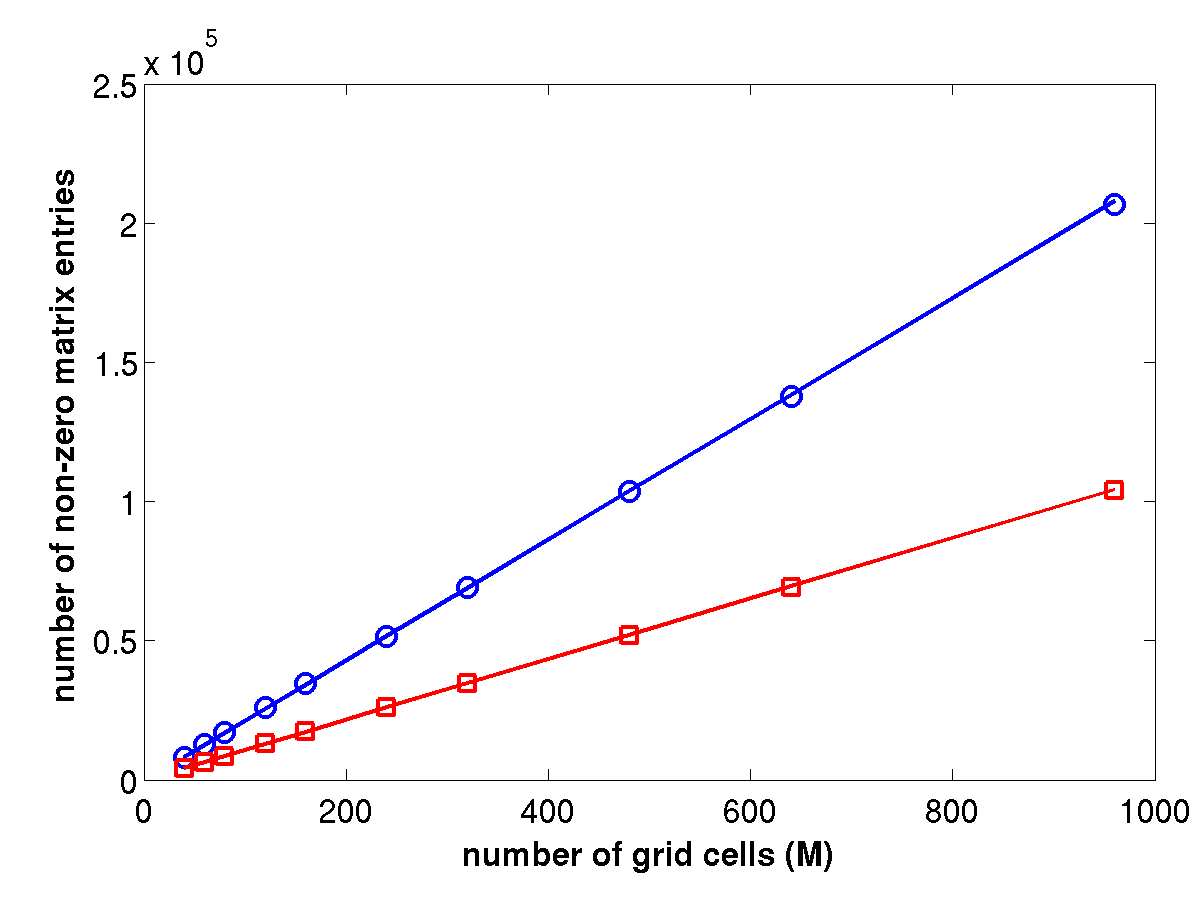}
		\caption{non-zero entries in $A$ (\textcolor{blue}{circles} - CP data with linear fit, \textcolor{red}{squares} - CACP data with linear fit)}
		\label{FIG:CP_vs_CACP_clover_nnz}
	\end{subfigure}
	\caption{comparing CACP method over CP method for the clover in $\mathbb{R}^2$}
	\label{FIG:CP_vs_CACP_clover}
\end{figure}

The clover shape, along with the CACP method solution, is seen in Figure \ref{FIG:clover_solution}.  The second-order accuracy of the CACP method is verified in Figure \ref{FIG:clover_convergence}.
As with the unit circle, Figure \ref{FIG:CP_vs_CACP_clover_errors} shows the error of the CACP method, normalized by the CP method error, for the clover.  As more grid nodes are added, the CACP method $L_2$ error is about $0.65$ of the CP method and the $L_\infty$ error about $0.8$ of the CP method.
The number of non-zero entries in $A$ is, again, shown for each method in Figure \ref{FIG:CP_vs_CACP_clover_nnz}.  The linear fit in the figure for the CP method results has a slope of approximately $217$, while the slope of the fit for the CACP method results is approximately $109$.  Thus, as with the circle, the CACP method produces less error than the CP method as $M$ increases, while the number of non-zero entries of $A$ grows at about half the rate.  Table \ref{TAB:condition_numbers_clover} in Appendix \ref{APP:condition_numbers} shows that the matrix condition numbers for both methods are again the same order of magnitude.

%%%%%%%%%%%%%%%%%%%%%
\subsection{Results in $\mathbb{R}^3$}

As with $\mathbb{R}^2$, the PDE (\ref{EQ:elliptic_SPDE}) is chosen in $\mathbb{R}^3$ following the lead of Chen and Macdonald \cite{chen_closest_2015} so that $\gamma(\theta,\psi) = \cos(3 \theta) \sin^3(\psi)(9\cos^2(\psi) -1)$ is the solution.  The PDE is solved on the domain $[$-$2,2] \times [$-$2,2] \times[$-$2,2]$, which is again discretized into $M+1$ grid nodes in each direction that are then enumerated.  The implementation of the CP method from \cite{chen_closest_2015} varies only slightly from that of $\mathbb{R}^2$:
\begin{align*}
	A \boldsymbol{\gamma} = \mathbf{m}, \text{ where } A = I - M \text{ and } M = E_1 L - \frac{6}{\Delta x^2}(I - E_3).
\end{align*}
The CACP method implementation still uses a variable-coefficient, centered-difference formula for (\ref{EQ:LB_relation3D}) with the \textit{interpolation} nodes, but the exact discretization may depend on the form of $T(\mbfx)$.  For the \textit{edge} nodes, the scaling factor $6/\Delta x^2$ is again carried over from the CP method.
\par
\begin{figure}[h]
	\center
	\begin{subfigure}[t]{0.49\textwidth}
		\center
		\includegraphics[width=\textwidth]{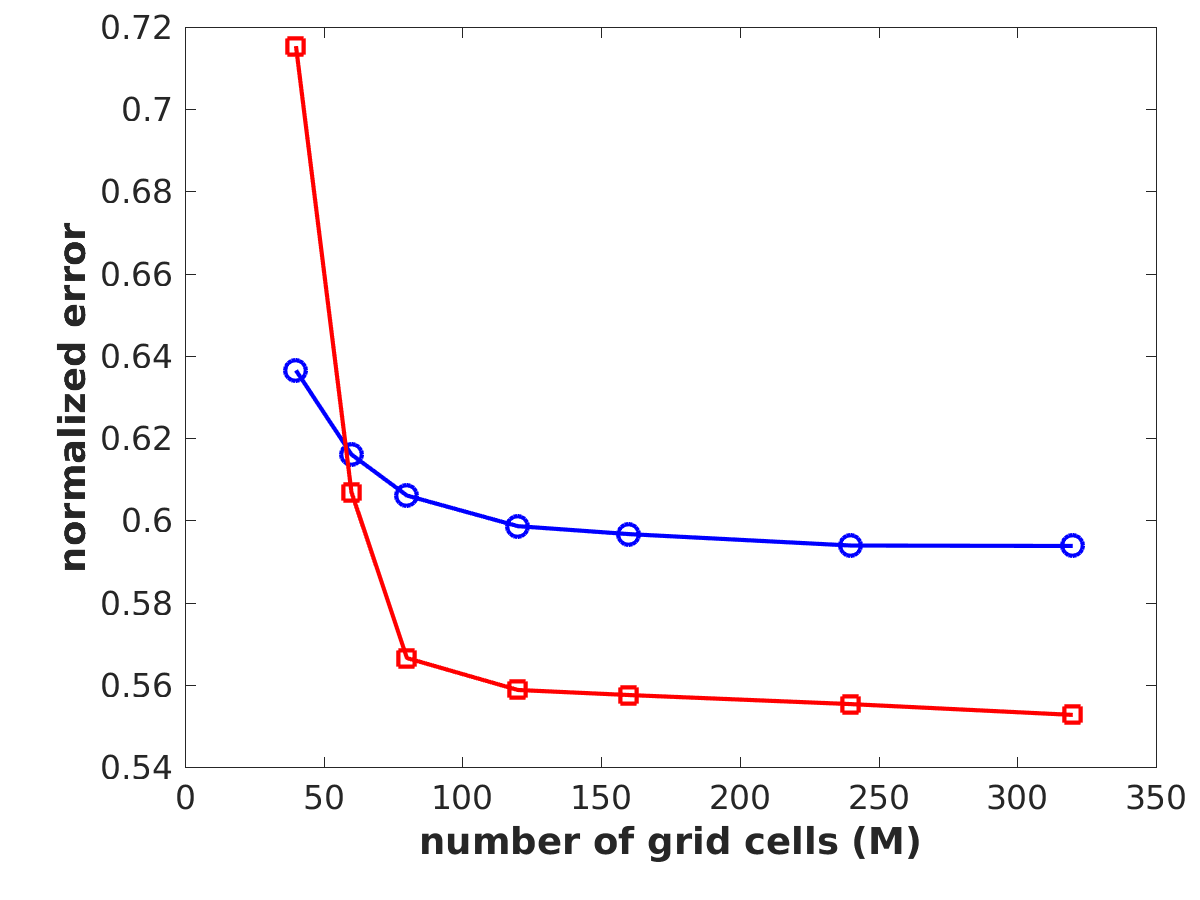}
		\caption{CACP error normalized by CP error (\textcolor{blue}{circles} - $L_2$ error, \textcolor{red}{squares} - $L_\infty$ error)}
		\label{FIG:CP_vs_CACP_sphere_errors}
	\end{subfigure}
	\begin{subfigure}[t]{0.49\textwidth}
		\center
		\includegraphics[width=\textwidth]{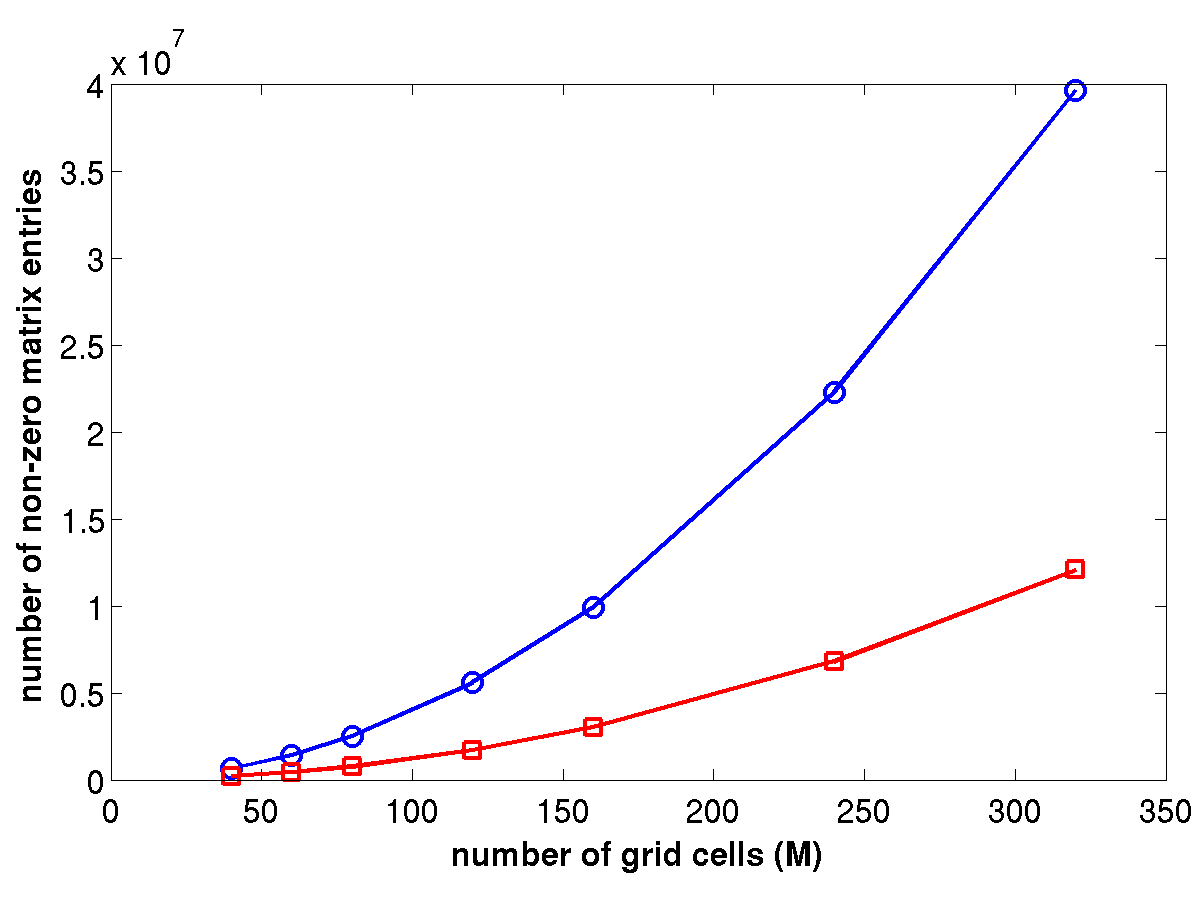}
		\caption{non-zero entries in $A$ (\textcolor{blue}{circles} - CP data with quadratic fit, \textcolor{red}{squares} - CACP data with quadratic fit)}
		\label{FIG:CP_vs_CACP_sphere_nnz}
	\end{subfigure}
	\caption{comparing CACP method to CP method for the unit sphere in $\mathbb{R}^3$}
	\label{FIG:CP_vs_CACP_sphere}
\end{figure}

Consider $\Gamma$ to be the unit sphere, so that $T(\mbfx)$ is simply the identity tensor.  Then $(\Delta_s \gamma)_{i,j,k} = (1 + \phi_{i,j,k})^2 (\Delta \gamma)_{i,j,k}$ where $(\Delta \gamma)_{i,j,k}$ is discretized with the same formula as the CP method.  Noting again that $\phi(\mbfx) = |\mbfx| - 1$, the second-order accuracy of  the CACP method is shown in Table \ref{TAB:CACP_errors_sphere}.
\begin{table}[h]
	\center
	\begin{tabular}{|l||
		S[round-mode=places,round-precision=4]|
		S[round-mode=places,round-precision=2]|
		S[round-mode=places,round-precision=4]|
		S[round-mode=places,round-precision=2]|}
		\hline
		& {$|\boldsymbol{\gamma}_\text{exact} - \boldsymbol{\gamma}_\text{num}|_2$} & {ratio} & {$|\boldsymbol{\gamma}_\text{exact} - \boldsymbol{\gamma}_\text{num}|_\infty$} & {ratio} \\
		\hline
		{$M = 40$} & 7.775448e-03 & {---} & 2.057475e-02 & {---} \\
		{$M = 80$} & 1.835869e-03 & 4.235295e+00 & 4.032997e-03 & 5.101603e+00 \\
		{$M = 160$} & 4.496780e-04 & 4.082630e+00 & 9.728364e-04 & 4.145606e+00 \\
		{$M = 320$} & 1.118339e-04 & 4.020945e+00 & 2.420839e-04 & 4.018591e+00 \\
		\hline
	\end{tabular}
	\caption{error values for the CACP method on the unit sphere in $\mathbb{R}^3$}
	\label{TAB:CACP_errors_sphere}
\end{table}
As with the previous two comparisons, Figure \ref{FIG:CP_vs_CACP_sphere} shows the normalized errors and compares the number of non-zero entries of $A$ between the CP and CACP methods.  As more grid nodes are added, the CACP method $L_2$ is about $0.6$ of the CP method and the $L_\infty$ about $0.56$ of the CP method.  Additionally, the number of non-zero entries in $A$ grows at a lower rate for the CACP than the CP method.  Here, the results are fit with $f(M) = aM^2 + b$, where $a$ is approximately $386$ for the CP method compared to $118$ for the CACP method.  Thus, the number of non-zero entries of $A$ for the CACP method grows at a third of the rate as for the CP method.  Again, the matrix condition numbers for both methods, shown in Table \ref{TAB:condition_numbers_sphere} in Appendix \ref{APP:condition_numbers}, are the same order of magnitude.

%%%%%%%%%%%%%%%%%%%%%
%%%%%%%%%%%%%%%%%%%%%
\section{Sample Application: Relaxing Vesicle}

With the accuracy of the CACP method verified and its improvement over the CP method observed, it is now applied to a problem where the accurate solving of an elliptic surface PDE is crucial for correct dynamics.  The bi-layer membrane of a lipid vesicle is often modeled as a bending resistant, inextensible interface between two fluids \cite{seifert_configurations_1997,vogl_effect_2014,kolahdouz_electrohydrodynamics_2015,rahimian_boundary_2015}.  For a vesicle of volume $\mathcal{V}$ and surface area $\mathcal{A}$, the reduced volume is defined as $\nu = \mathcal{V} / [(4\pi/3) R_0^3]$, where $R_0$ is the radius of a sphere with a surface area equal to that of $\mathcal{A}$: $R_0 = (\mathcal{A} / 4\pi)^{1/2}$.  This value determines what behavior a vesicle exhibits under flow \cite{kantsler_orientation_2005,salac_reynolds_2012} and what equilibrium shape the vesicle seeks at rest \cite{seifert_shape_1991,seifert_configurations_1997}.  The latter is demonstrated here, with the inextensibility of the vesicle enforced via an elliptic surface PDE.
\par
The bending resistance and inextensibility of the vesicle result in a membrane energy  of the form $\int_\Gamma [(b/2) H(\mbfx,t)^2 + \gamma(\mbfx,t)] dS$, where $b$ is a material bending parameter, $H$ is the total curvature $\kappa + h$, and $\gamma$ is a surface tension that enforces the inextensibility, as discussed by \cite{seifert_fluid_1999}.  The surrounding fluid is assumed to be incompressible and Newtonian, contained in a domain $\Omega$.  Imposing free-slip conditions at $\partial \Omega$ results in the following non-dimensionalized Navier-Stokes equation for the fluid velocity $\mbfu(\mbfx,t)$:
\begin{align*}
	\text{Re} (\mbfu_t + \nabla \mbfu . \mbfu) = -\nabla p + \Delta \mbfu + \delta(\phi) \mathbf{f}_\Gamma, \quad \mbfu |_{t=0} = \mathbf{0}, \quad \nabla \cdot \mbfu = 0, \\
	\mbfu |_{\partial \Omega} \cdot \mbfn_\Omega = 0, \quad (I - \mbfn_\Omega \otimes \mbfn_\Omega)[(\nabla \mbfu + \nabla^T \mbfu)|_\Omega . \mbfn_\Omega] = \mathbf{0},\\
	\text{and} \quad \mathbf{f}_\Gamma = \text{Be} ( \Delta_s H - 2KH + \frac{1}{2}H^3)\mbfn + \nabla_s \gamma - H\gamma\mbfn,
\end{align*}
where $\mbfn_\Omega$ is the outward pointing normal for $\Omega$.  The membrane force $\mathbf{f}_\Gamma$ is obtained via a variational derivative of the energy.  Note the standard Reynolds number Re $= \rho u / (\mu/L)$, where $\rho$ and $\mu$ are the density and viscosity of the surrounding fluid, respectively, while $u$ and $L$ are a characteristic speed and length scale, respectively.  The other dimensionless parameter is the bending number Be, defined as $b / (\mu u L^2)$.  Additionally, $\phi(\mbfx,t)$ is the signed-distance level set function for $\Gamma$, and $K$ is the Gaussian curvature $\kappa h$.
\par
A final equation is needed to determine $\gamma(\mbfx,t)$ that enforces the inextensibility of the membrane.  Denote $\tilde{\mbfu}(\sigma_1,\sigma_2,t) = \mbfu(\mbfr(\sigma_1,\sigma_2,t),t)$ and note that inextensibility requires $\mbfr_{\sigma_i}  g^{ij} \tilde{\mbfu}_{\sigma_j} = 0$ from \cite{seifert_fluid_1999}.  In order to simplify this expression, the vesicle is assumed to always be axisymmetric and, as a result, that the fluid flow is also axisymmetric.  Using the cylindrical coordinate system from Section \ref{SEC:improvement_R3_axis}, where $\mbfr(\sigma,\theta,t) = x(\sigma,t)\mathbf{e}_x(\theta) + y(\sigma,t)\mathbf{e}_y$, the inextensibility requirement simplifies to
\begin{align}
	\frac{1}{x}\frac{1}{|\mbfr_\sigma|}\big(x (\tilde{\mbfu} \cdot \mbft^\sigma)\big)_\sigma + H (\tilde{\mbfu} \cdot \mbfn) = 0, \label{EQ:inextensibility}
\end{align}
with additional details found in Appendix \ref{APP:inextensibility}.  Similar to how the pressure $p$ is determined from the incompressibility condition $\nabla \cdot \mbfu = 0$, the tension is eventually determined by (\ref{EQ:inextensibility}).

%%%%%%%%%%%%%%%%%%%%%
\subsection{Numerical Approach}

Given that the reduced volume $\nu$ plays an important role in determining the equilibrium shape, it is important to conserve the volume of the vesicle along with the surface area.  As such, the level set approach of Vogl \cite{vogl_curvature-augmented_2016} is used for its ability to minimize mass loss.  Note that in order to advance $\phi^n := \phi(\mbfx,t_n)$ to $\phi^{n+1} := \phi(\mbfx,t_{n+1})$, that approach requires $\mbfu^{n+1/2}$.  Discrete-divergence free velocity fields are obtained by decomposing the right-hand side of the Navier-Stokes equation as
\begin{align}
	-\text{Re} \nabla \mbfu . \mbfu + \Delta \mbfu = \nabla \times \Psi_f  + \nabla p_f, \label{EQ:decomp_f}\\
	\delta(\phi)\text{Be}(\Delta_s H - 2KH + \frac{1}{2}H^3) \mbfn = \nabla \times \Psi_b + \nabla p_b, \label{EQ:decomp_b}\\
	\delta(\phi)(\gamma_s \mbft^\sigma - H \gamma \mbfn) = \nabla \times \Psi_\gamma + \nabla p_\gamma, \label{EQ:decomp_g}
\end{align}
where $\Psi_f = \psi_f\mathbf{e}_\theta$, $\Psi_b = \psi_b \mathbf{e}_\theta$, and $\Psi_\gamma = \psi_\gamma \mathbf{e}_\theta$.  Given that $\mbfu$ is divergence-free, implying that $\mbfu_t$ is also divergence-free, the values for $\mbfu^{n+1/2}$ are obtained via first-order integration in time:
\begin{align*}
	\mbfu^{n+1/2} = \mbfu^n + \frac{\Delta t}{2\text{Re}}\big(\nabla \times \Psi_f^n  + \nabla \times \Psi_b^{n+1/2} + \nabla \times \Psi_\gamma^{n+1/2}\big).
\end{align*}
The stream function $\psi_f^n$ is obtained by solving the Poisson equation that results from taking a cross product of (\ref{EQ:decomp_f}), followed by a dot product with $\mathbf{e}_\theta$.  A similar process obtains $\psi_b^{n+1/2}$,
where interface quantities are extrapolated to obtain their respective values at $t_{n+1/2}$ ($\phi^{n+1/2} = 1.5\phi^n - 0.5\phi^{n-1}$, $\kappa^{n+1} = 1.5 \kappa^n - 0.5\kappa^{n-1}$, etc).
\par
Obtaining the stream function $\psi_\gamma^{n+1/2}$ requires more work, as the tension $\gamma^{n+1/2}$ must first be determined.  Applying the inextensibility condition (\ref{EQ:inextensibility}) to $\tilde{\mbfu}^{n+1/2}$ eventually leads to
\begin{align*}
	H^2 \gamma - \Delta_s \gamma = &-\frac{1}{\delta(0)}\left [\frac{1}{x}\big( x (\tilde{\nabla p}_\gamma \cdot \mbft^\sigma ) \big)_s + H (\tilde{\nabla p}_\gamma \cdot \mbfn)\right] \\
&+\frac{2\text{Re}}{\delta(0)\Delta t} \left[ \frac{1}{x}\big( x (\tilde{\mbfu}^* \cdot \mbft^\sigma) \big)_s + H (\tilde{\mbfu}^* \cdot \mbfn) \right],
\end{align*}
where $\mbfu^* = \mbfu^n + 0.5(\Delta t / \text{Re}) [ \nabla \times \Psi_f^n  + \nabla \times \Psi_b^{n+1/2} ]$.  Note the equation for the tension is now in the form of (\ref{EQ:elliptic_SPDE}), so the CACP method is applied.  The pressure term $p_\gamma^{n+1/2}$ is also an unknown at this point, thus the surface PDE above is coupled with the Poisson equation that results in taking the divergence of (\ref{EQ:decomp_g}) and solved all at once.  The stream function $\psi^{n+1/2}_\gamma$ is then found in a similar fashion to the other stream functions.  Additional details, including grid discretization and finite-difference stencils, are found in Appendix \ref{APP:numerical_details}.
\par
With $\mbfu^{n+1/2}$ now determined, the level set is advanced, as in \cite{vogl_curvature-augmented_2016}, from $t_n$ to $t_{n+1}$.  As the interface quantities at $t_{n+1}$ are now available, the quantities at $t_{n+1/2}$ are updated using interpolation instead of extrapolation ($\phi^{n+1/2} = 0.5\phi^n + 0.5\phi^{n+1}$, etc).  The stream functions are then all  recalculated in one Poisson solve, using  $\psi = \psi_f + \psi_b + \psi_\gamma$.  The velocity is then advanced in a Runge-Kutta fashion:
\begin{align*}
	\mbfu^{n+1} = \mbfu^n + \frac{\Delta t}{\text{Re}}\big( \nabla \times \Psi^{n+1/2}\big).
\end{align*}
While the number of Poisson solves for the stream functions can likely be decreased with modification to this approach, the focus here is on the accurate solving of the tension surface PDE.

%%%%%%%%%%%%%%%%%%%%%
\subsection{Results}

With the numerical method in place, vesicles of various initial shapes are simulated as they relax to their equilibrium shapes.  Referring to work of Seifert, Berndl, and Lipowsky \cite{seifert_shape_1991}, these shapes depend on the reduced volume $\nu$.  Oblate shapes are possible from $\nu = 1$ down until about $\nu = 0.5$, where they self-intersect.  Stomatocyte shapes are found for $\nu$ less than $0.66$.  Defining a computational domain of $(x,y) \in [0,1/2] \times [0,1]$ for $\mbfx = x \mathbf{e}_x(\theta) + y\mathbf{e}_y$, this motivates initial shape level sets of the form $l(x,y) = (x/0.35)^2 + ([y-(0.5 + c x^2)]/0.1)^2 - 1$.  Here, $\nu$ decreases from $0.64$ as $c$ increases from $0$.  The relative errors in the surface area, volume, and $\nu$ are monitored as the vesicles evolve, using numerical integration of the regularized delta and heaviside functions \cite{vogl_curvature-augmented_2016}.
\par
\begin{figure}[h]
	\center
	\begin{subfigure}{0.49\textwidth}
		\center
		\includegraphics[width=0.49\textwidth]{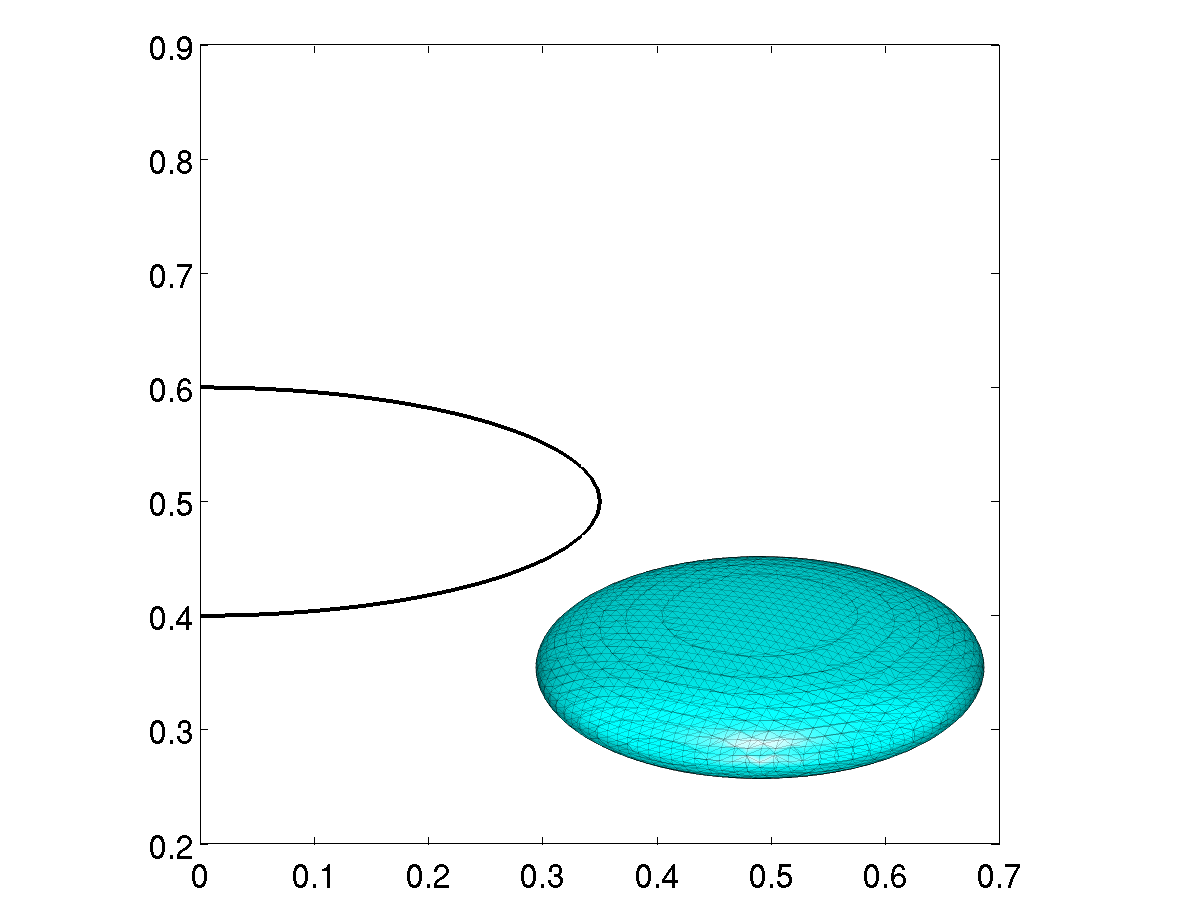}
		\includegraphics[width=0.49\textwidth]{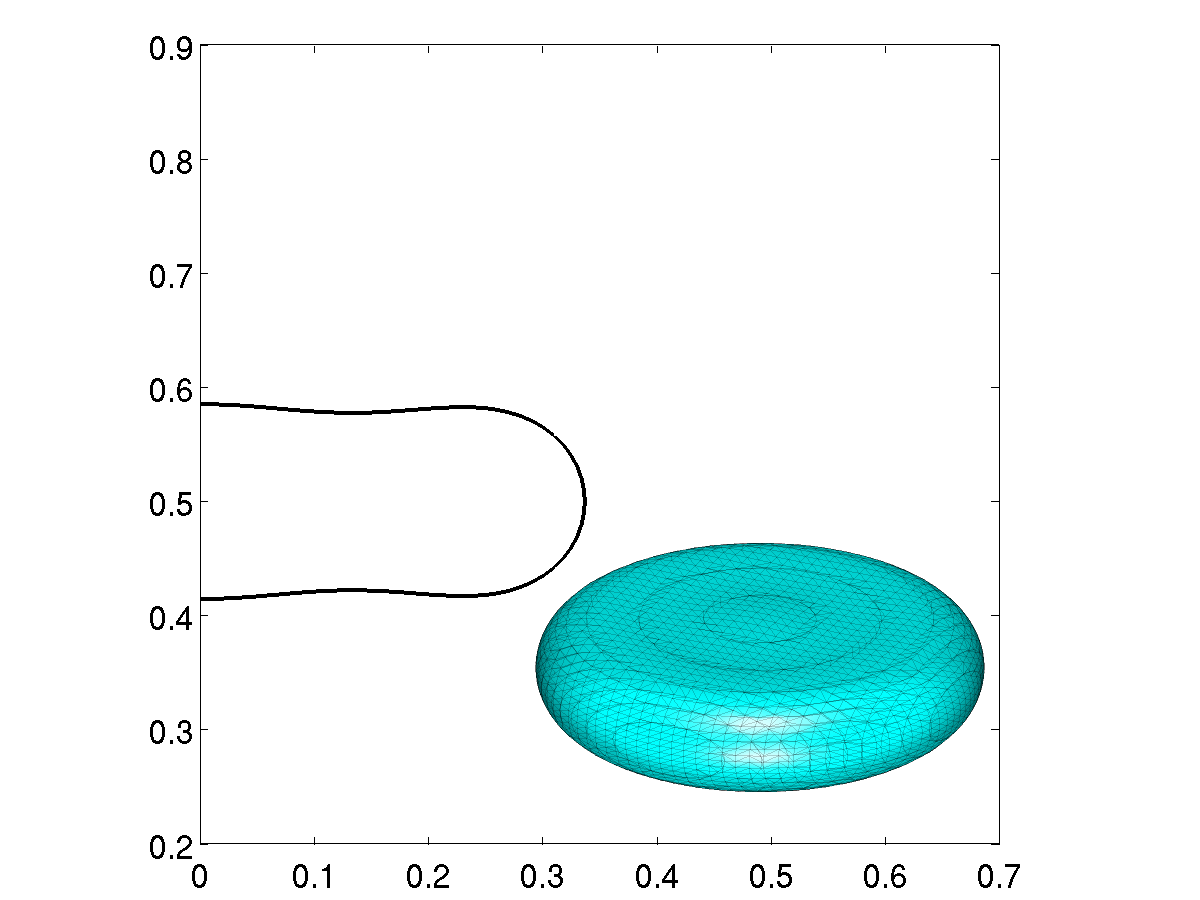}\\
		\includegraphics[width=0.49\textwidth]{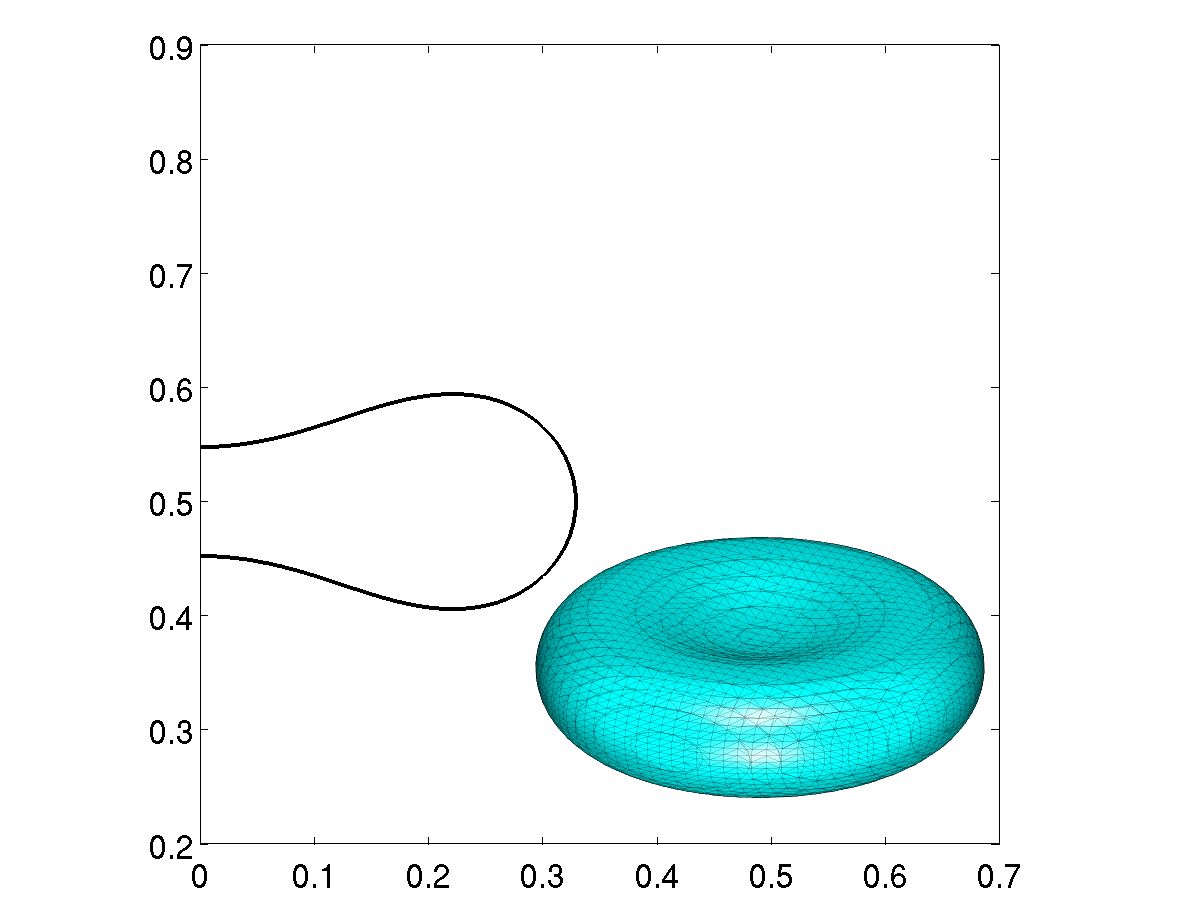}
		\includegraphics[width=0.49\textwidth]{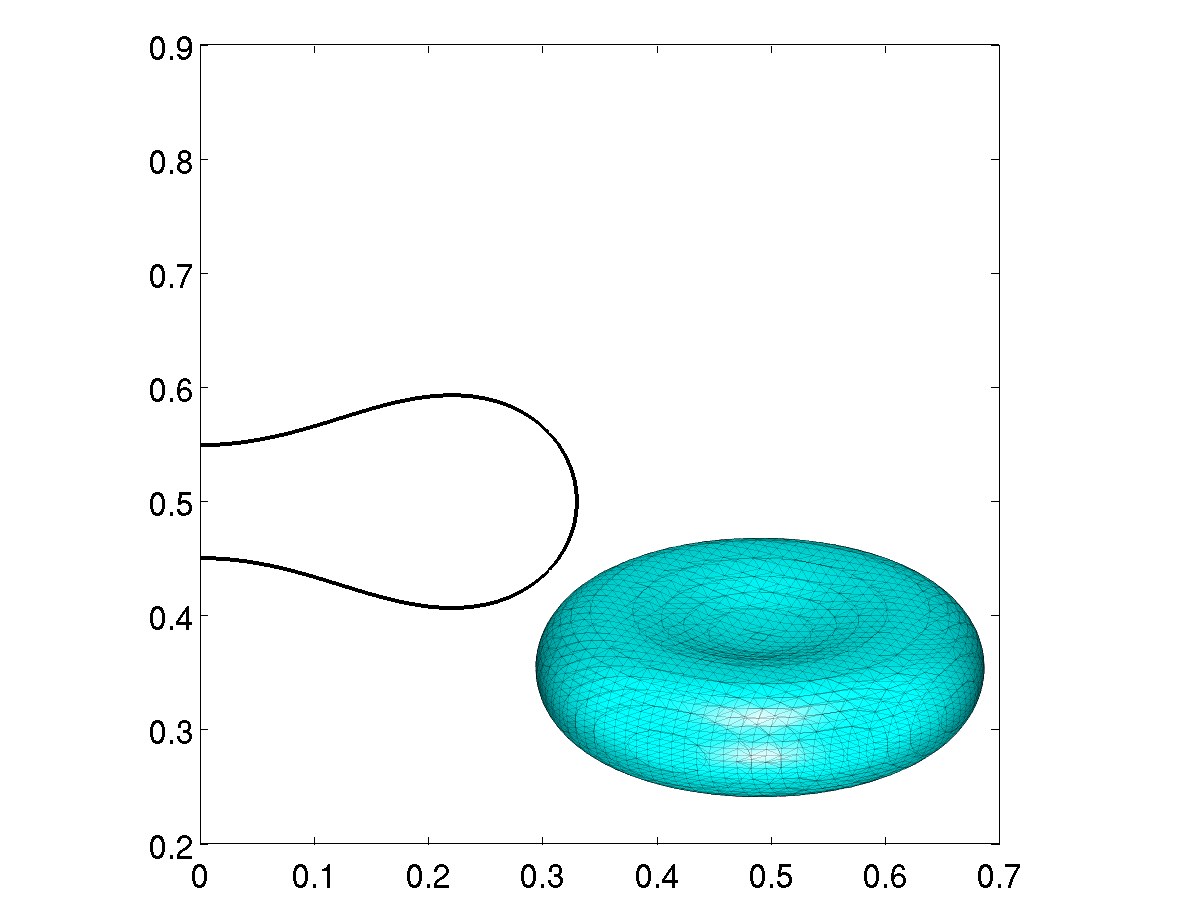}
		\caption{time captures of shapes ($t$ from left-to-right, top-to-bottom: 0.0, 2.5E-3, 5.0E-3, 5.0E-2)}
		\label{FIG:vesicle_c00_shapes}
	\end{subfigure}
	\begin{subfigure}{0.49\textwidth}
		\center
		\includegraphics[width=\textwidth]{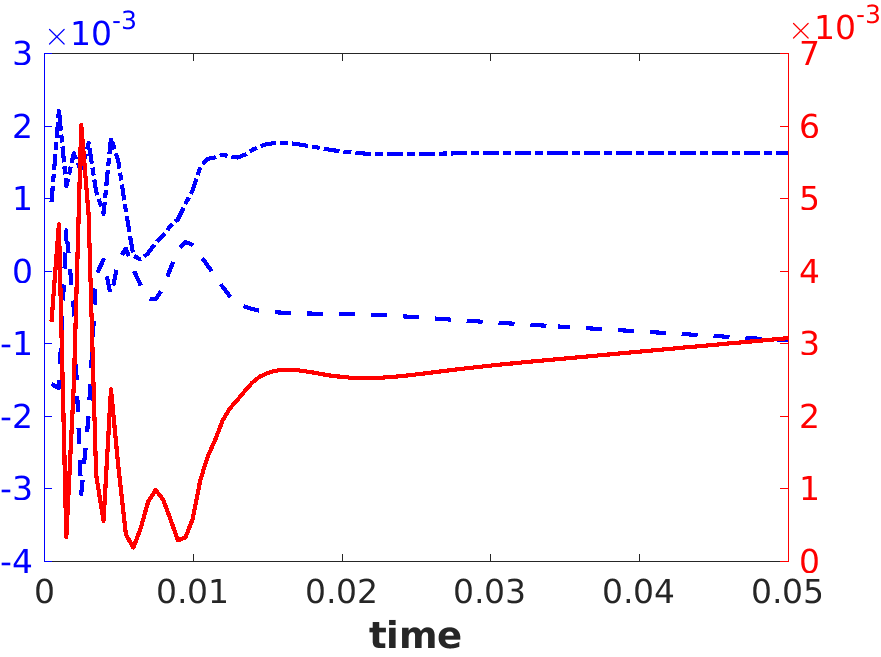}
		\caption{relative errors (left scale: \textcolor{blue}{dashed} - surface area, \textcolor{blue}{dot-dashed} - volume; right scale: \textcolor{red}{solid} - reduced volume)}
		\label{FIG:vesicle_c00_errors}
	\end{subfigure}
	\caption{initial vesicle shape of $c=0.0$ ($\nu = 0.64$) relaxing to an oblate shape}
\end{figure}

To start, the ellipsoidal vesicle $(c = 0.0)$ is observed relaxing to its equilibrium shape.  This is done on a $50 \times 100$ cell grid with a timestep of $5.0$E-$6$ for $t \in [0,0.05]$ and Re=Be=1.  In order to preserve the signed-distance function, a full-reinitialization is performed every $200$ iterations.  Figure \ref{FIG:vesicle_c00_shapes} shows the progression of the vesicle to the equilibrium oblate shape, with the high curvature regions rounding out first and the center thinning after.  The latter of these dynamics is a result of the inextensibility constraint, without which the vesicle would become a sphere.  Figure \ref{FIG:vesicle_c00_errors} shows that the inextensibility and incompressibility are maintained with a relative error less than $0.4 \%$ and the reduced volume $\nu$ held to within $0.7 \%$ relative error.
\par
\begin{figure}[h]
	\center
	\begin{subfigure}{0.49\textwidth}
		\center
		\includegraphics[width=0.49\textwidth]{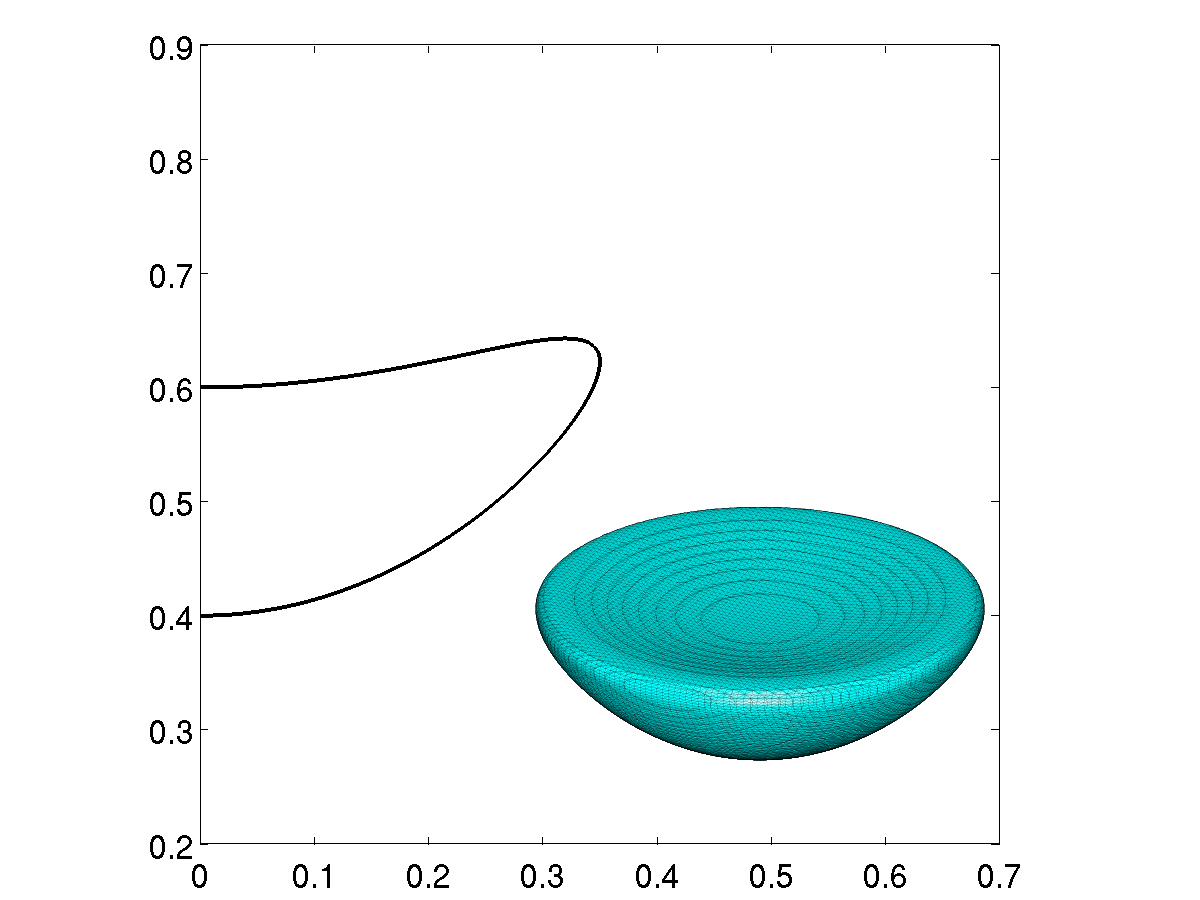}
		\includegraphics[width=0.49\textwidth]{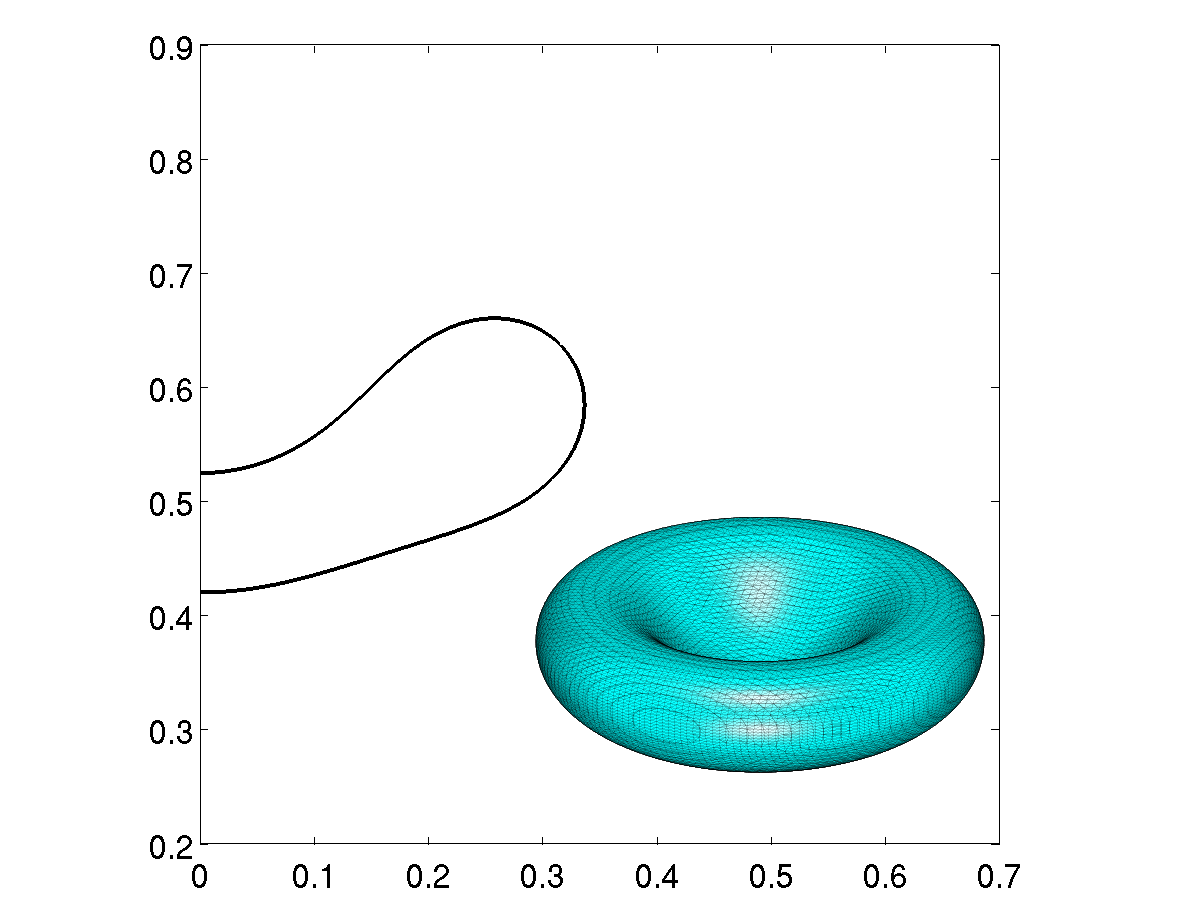}\\
		\includegraphics[width=0.49\textwidth]{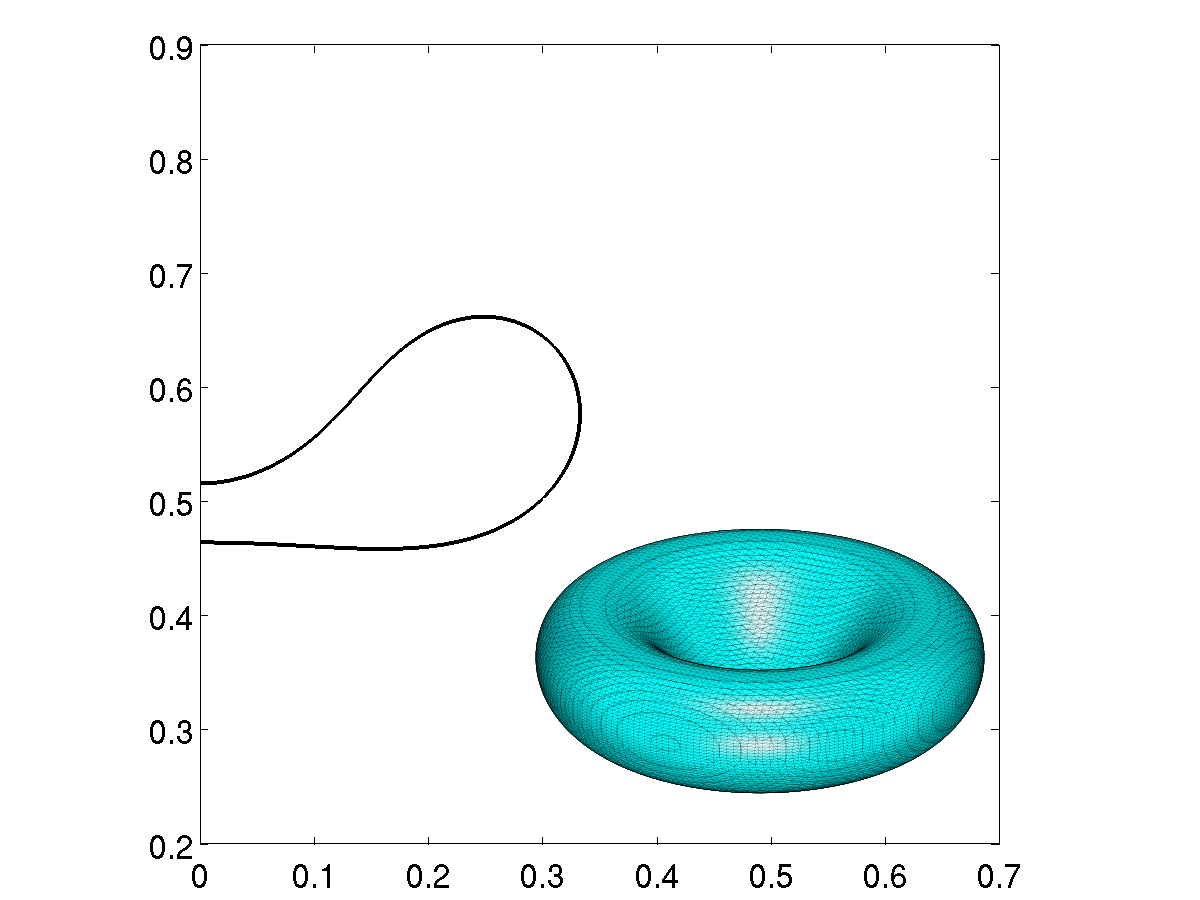}
		\includegraphics[width=0.49\textwidth]{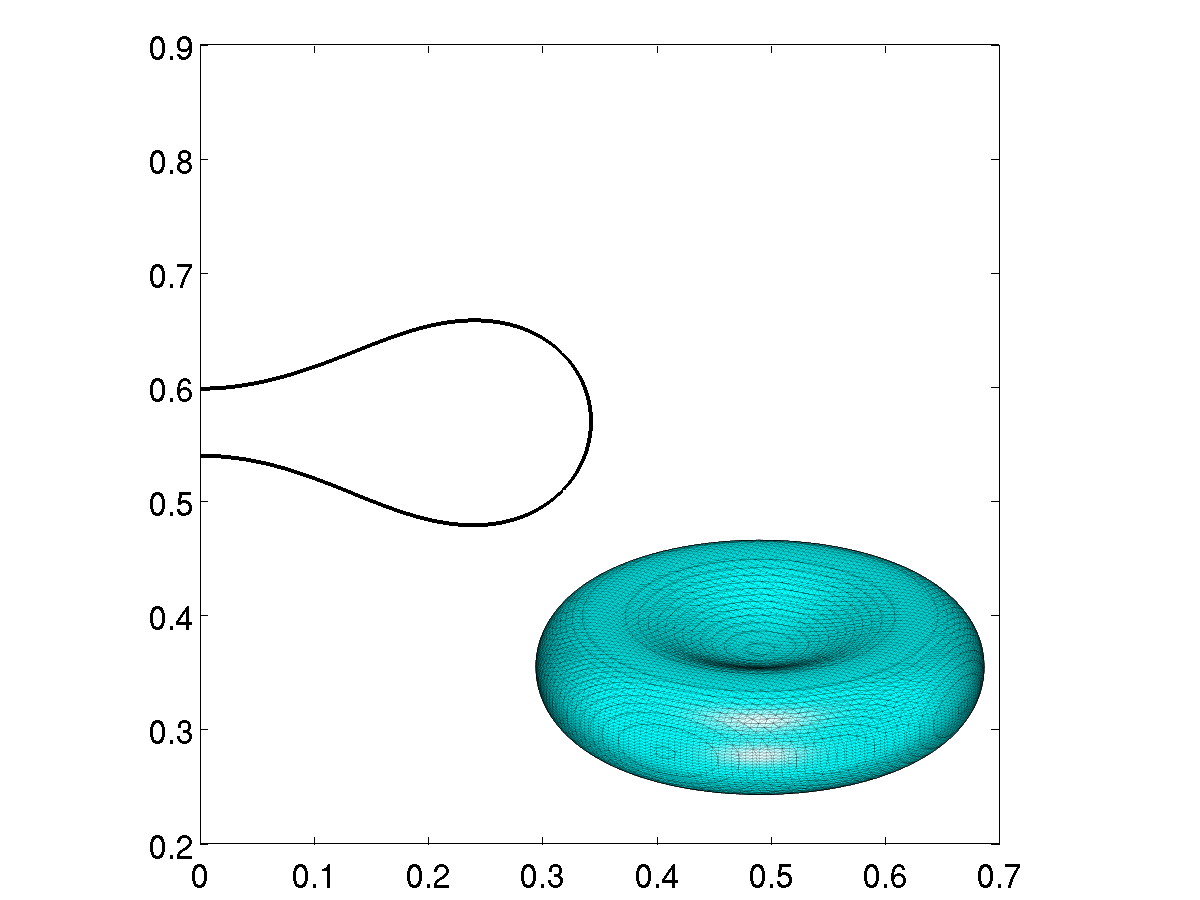}
		\caption{time captures of shapes ($t$ from left-to-right, top-to-bottom: 0.0, 5.0E-3, 1.0E-2, 5.0E-2)}
		\label{FIG:vesicle_c10_shapes}
	\end{subfigure}
	\begin{subfigure}{0.49\textwidth}
		\center
		\includegraphics[width=\textwidth]{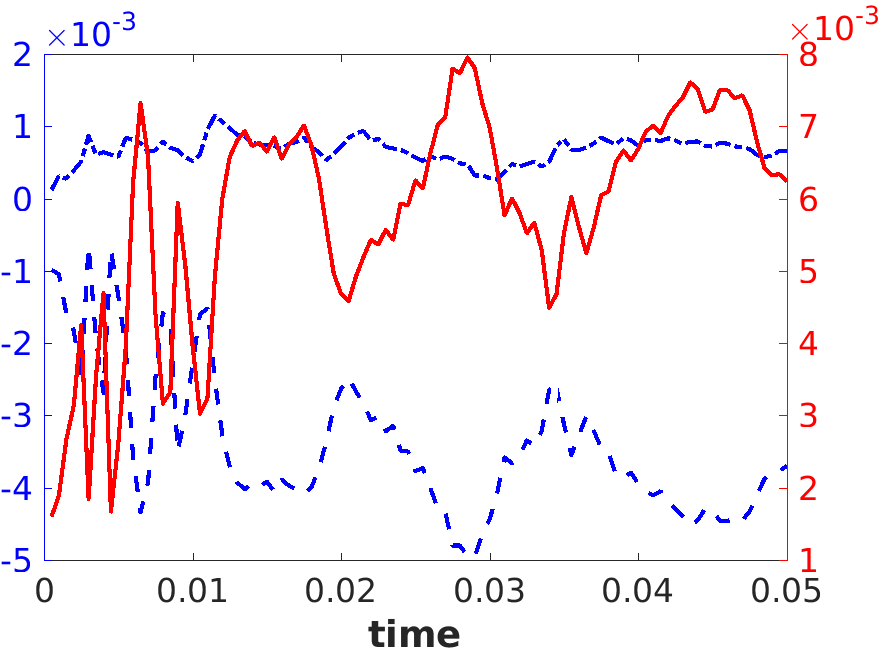}
		\caption{relative errors (left scale: \textcolor{blue}{dashed} - surface area, \textcolor{blue}{dot-dashed} - volume; right scale: \textcolor{red}{solid} - reduced volume)}
		\label{FIG:vesicle_c10_errors}
	\end{subfigure}
	\caption{initial vesicle shape of $c=0.1$ ($\nu = 0.58$) relaxing to an oblate shape}
\end{figure}

The up-down symmetry of the initial shape is broken next, with $c = 1.0$.  This results in a reduced volume $\nu = 0.58$, so both oblate and stomatocyte equilibrium shapes are possible. The vesicle is simulated on a $100 \times 200$ cell grid with a timestep of $1.25$E-$6$ for $t \in [0,0.05]$, where the extra resolution is needed to resolve the higher curvature regions of the initial shape.  The Reynolds number, bending number, and full reinitialization interval remain the same as with $c=0.0$.  While both oblate and stomatocyte shapes are locally stable for $\nu = 0.58$, the oblate shape has the lower total energy \cite{seifert_shape_1991}.  This is reflected in Figure  \ref{FIG:vesicle_c10_shapes}, where the vesicle takes a slightly different progression to the oblate shape.  Here, the thinning of the center first furthers the up-down asymmetry, but eventually the vesicle returns to being symmetric.  The relative errors are again found in Figure \ref{FIG:vesicle_c10_errors}, showing the surface area and volume maintained to within $0.5\%$ and the reduced volume $\nu$ held to within $0.8 \%$ relative error.
\par
\begin{figure}[h]
	\center
	\begin{subfigure}{0.49\textwidth}
		\center
		\includegraphics[width=0.49\textwidth]{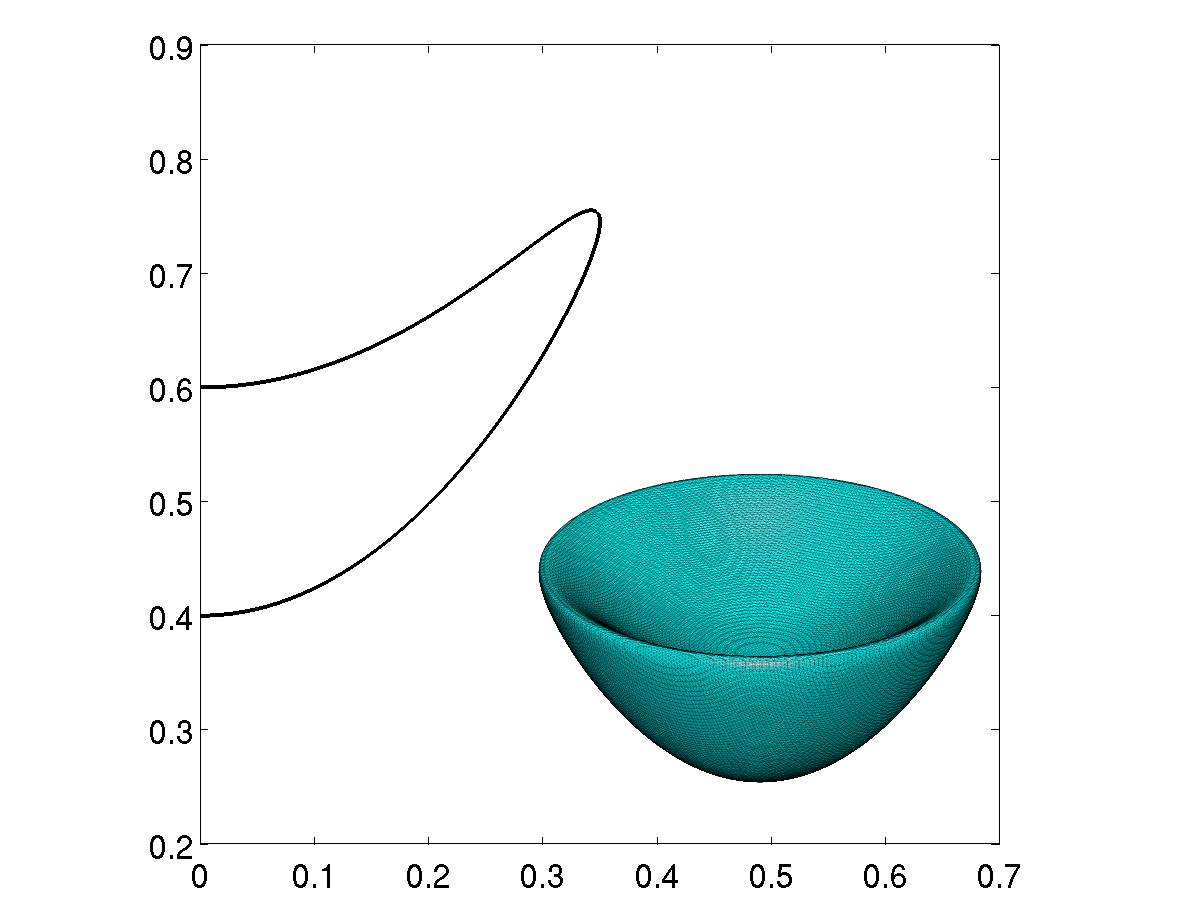}
		\includegraphics[width=0.49\textwidth]{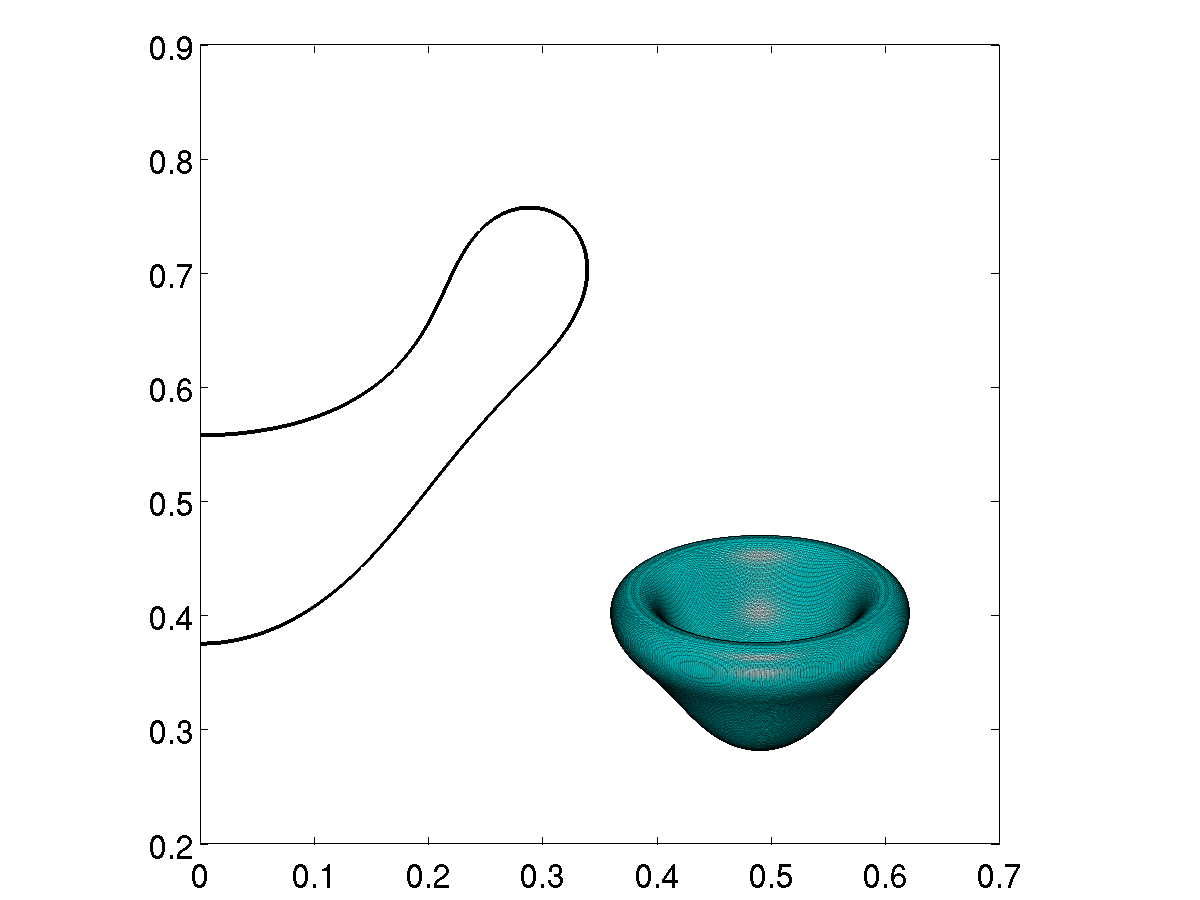}\\
		\includegraphics[width=0.49\textwidth]{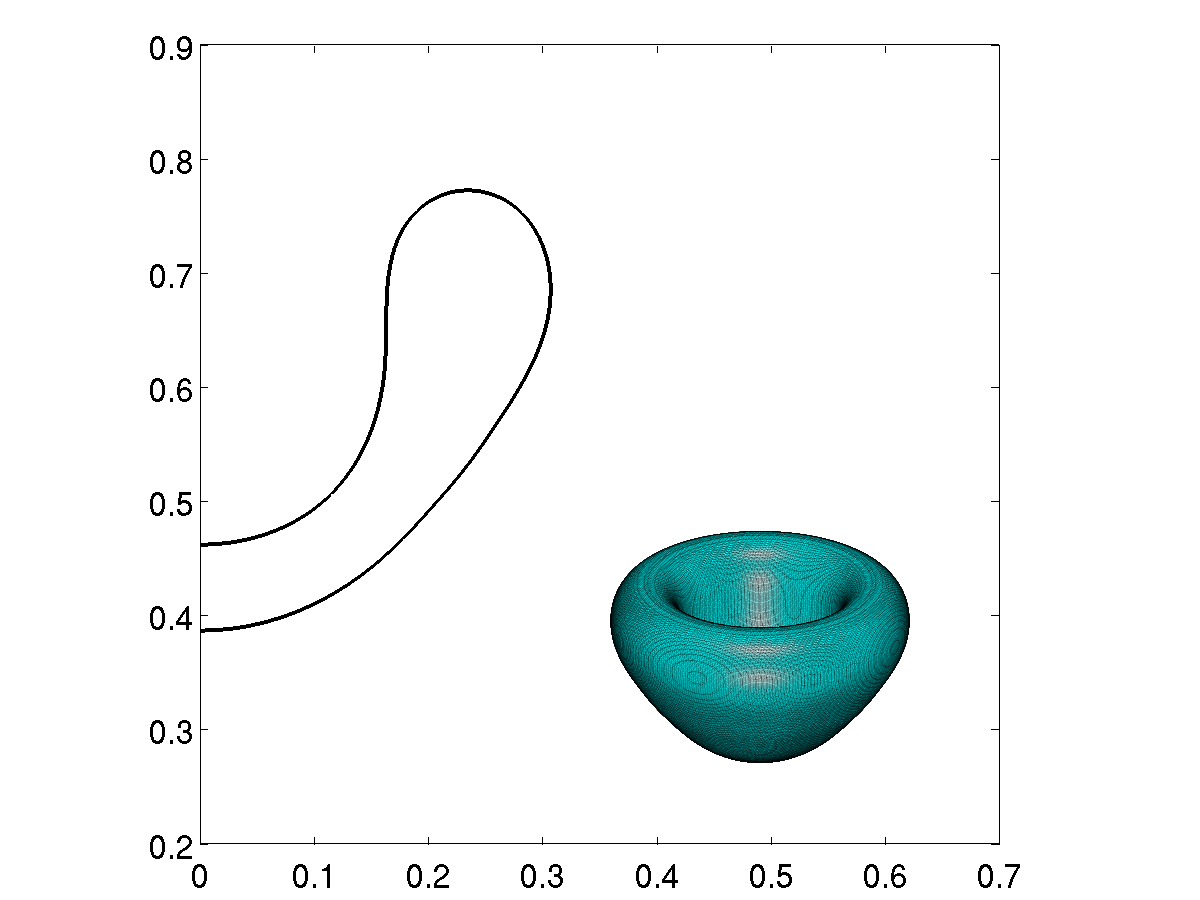}
		\includegraphics[width=0.49\textwidth]{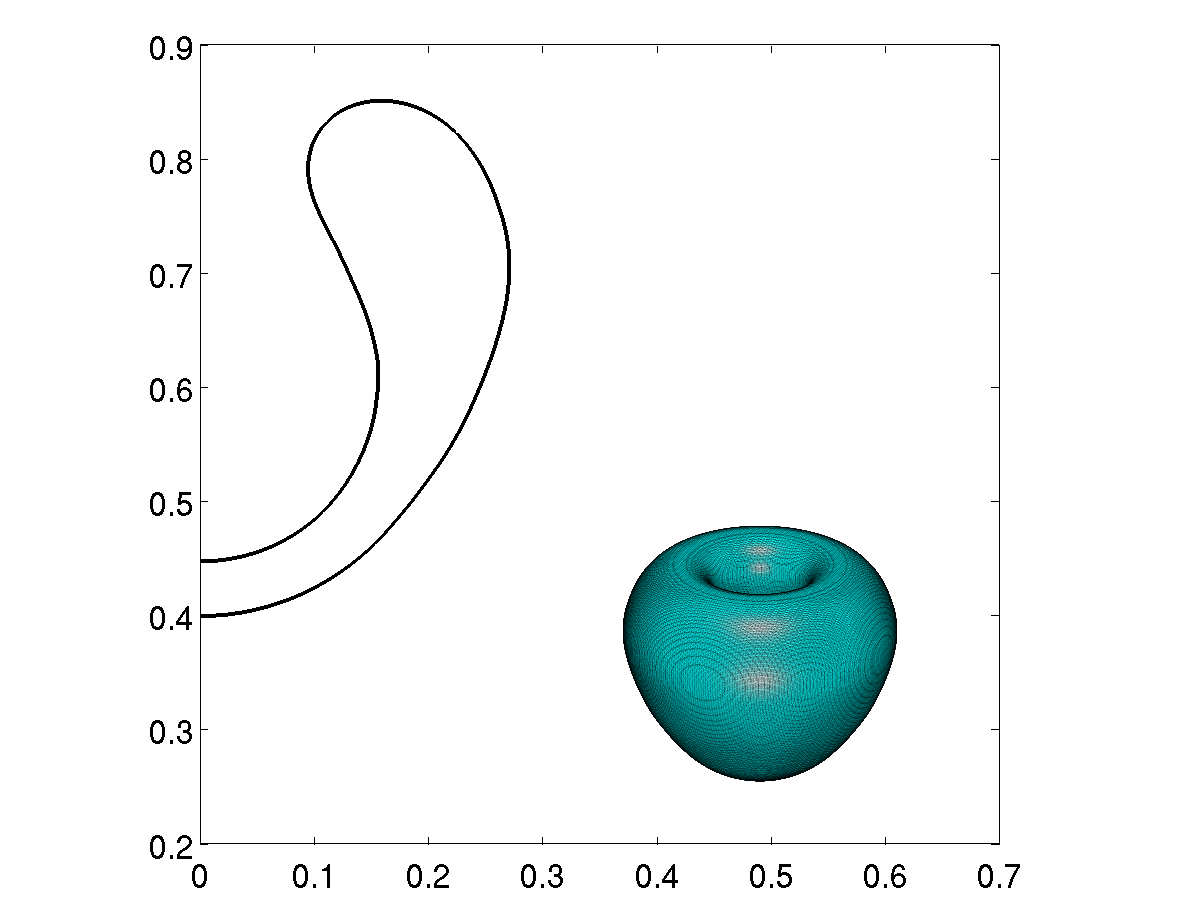}
		\caption{time captures of shapes ($t$ from left-to-right, top-to-bottom: 0.0, 5.0E-3, 2.0E-2, 5.0E-2)}
		\label{FIG:vesicle_c20_shapes}
	\end{subfigure}
	\begin{subfigure}{0.49\textwidth}
		\center
		\includegraphics[width=\textwidth]{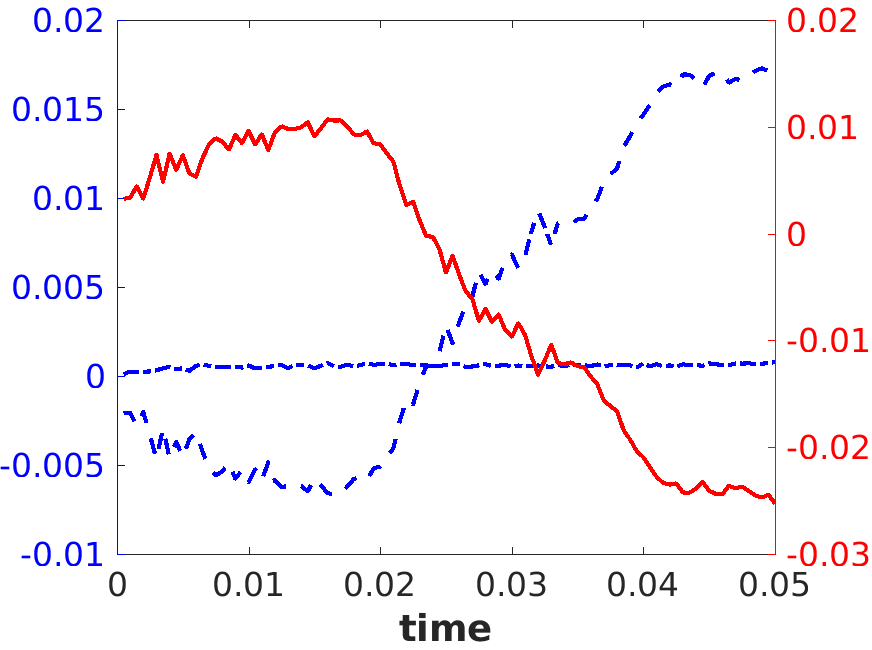}
		\caption{relative errors (left scale: \textcolor{blue}{dashed} - surface area, \textcolor{blue}{dot-dashed} - volume; right scale: \textcolor{red}{solid} - reduced volume)}
		\label{FIG:vesicle_c20_errors}
	\end{subfigure}
	\caption{initial vesicle shape of $c=0.2$ ($\nu = 0.45$) relaxing to a stomatocyte shape}
\end{figure}

The final shape is with $c = 2.0$, which further breaks the up-down symmetry.  This value of $c$ results in $\nu = 0.45$, where the oblate equilibrium shape no longer exists.  The vesicle is simulated on a $200 \times 400$ cell grid with a timestep of $3.125$E-$7$ for $t \in [0,0.05]$.  The Reynolds number, bending number, and full reinitialization interval remain unchanged.  It is only now that the vesicle relaxes to the stomatocyte shape.  As seen in Figure  \ref{FIG:vesicle_c20_shapes}, the thinning of the vesicle furthers the up-down asymmetry to a point where the vesicle continues away from a symmetric shape and into an equilibrium stomatocyte shape.  This is important when considering a vesicle as prototype model for a red blood cell: without the stomatocyte shape, the vesicle self-intersects for this reduced volume.  Figure \ref{FIG:vesicle_c20_errors} show the relative surface area and volume errors that, while higher than the $c=0.0$ and $c=1.0$ shapes due to the higher initial shape curvature, remain within $1.5\%$.  The reduced volume $\nu$ is held to within $2.0\%$ relative error.

%%%%%%%%%%%%%%%%%%%%%
%%%%%%%%%%%%%%%%%%%%%
\section{Conclusion}

The Closest Point (CP) method provides a way to solve surface PDEs, like (\ref{EQ:elliptic_SPDE}), using standard finite-difference and interpolation stencils without having to worry about parameterizing the surface.  This is done by solving an alternative PDE where the surface operators are embedded into the Cartesian domain (\ref{EQ:LB_relation}).  The Curvature-Augmented Closest Point (CACP) method improves on this by deriving generalized operator embeddings that are valid both on and around the surface.  These are obtained for the surface gradient and Laplace-Beltrami operators on both one-dimensional surfaces in $\mathbb{R}^2$ (\ref{EQ:LB_relation2D}) and two-dimensional surfaces in $\mathbb{R}^3$ (\ref{EQ:LB_relation3D}), with specific embeddings found for a sphere (\ref{EQ:LB_relation3D_sphere}) and an axisymmetric surface (\ref{EQ:LB_relation3D_axis}).  The CACP is compared against the CP method for test cases, and then applied to a realistic model for inextensible vesicle evolution.
\par
For a circle in $\mathbb{R}^2$, the CACP method results have an $L_2$ error that is about $0.62$ of the CP method results and an $L_\infty$ error that is about $0.68$ of the CP method.  The resulting linear system involved in obtaining the CACP method results is also more sparse than that for the CP method results.  The number of non-zero entries in both the CP and CACP method matrices grow linearly with the number of grid cells in one direction; however, the growth rate for the CACP method is about half  that of the CP method.  For a clover shape in $\mathbb{R}^2$, the CACP results have $L_2$ and $L_\infty$ errors that are $0.65$ and $0.8$ of the CP method results, respectively.  The number of non-zero entries for the CACP method still grows linearly at half  the rate of the CP method.  For a sphere in $\mathbb{R}^3$, the CACP method results have errors that are $0.6$ and $0.56$ of the CP method.  The number of non-zero entries for both the CP and CACP methods grows quadratically with the number of grid cells in one direction, but the growth rate for the CACP method is a third of that for the CP method.  In all test cases, the matrix condition numbers from the CACP method are the same order of magnitude as those from the CP method, indicating the increased sparsity does not come with an increased solution cost.
\par
With the CACP method verified, it is applied to modeling a vesicle relaxing in a fluid.  Without surface inextensibility, the surfaces would all relax to spheres to minimize the corresponding bending energy.  Enforcing the inextensibility is accomplished by a surface tension governed by an elliptic surface PDE of the form of (\ref{EQ:elliptic_SPDE}).  Additionally, the forcing term from the surface on the fluid involves the surface gradient of tension and the Laplace-Beltrami operator acting on the total curvature, making this problem very suitable for the CACP method.  The resulting solutions conserve the surface area to less than $0.4 \%$, $0.5 \%$, and $1.5 \%$ error for the various initial shapes and grid resolutions chosen.  The resulting evolution of the vesicle shapes show both the oblate and stomatocyte equilibrium shapes found by Seifert \cite{seifert_shape_1991}, indicating the successful application of the CACP method.
\par
Future work involves investigating the performance of this method on arbitrary surfaces in $\mathbb{R}^3$.  The operator embeddings (\ref{EQ:LB_relation3D}) are defined so long as the principal directions and curvatures are attainable.  Given a sufficiently smooth level set $l(\mbfx)$, the shape tensor is given by $\nabla [\nabla l(\mbfx) / |\nabla l(\mbfx)|]. \mbfr_{\sigma_j} \cdot \mbfr_{\sigma_i}$.  Thus, the principal directions and curvatures can be found at $\mbfx$ by either  identifying eigenvectors and eigenvalues of $\nabla [\nabla l(\mbfx) / |\nabla l(\mbfx)|]$ or finding extrema of $\nabla [\nabla l(\mbfx) / |\nabla l(\mbfx)|].\mathbf{v}\cdot \mathbf{v}$, for $|\mathbf{v}| = 1$.  The CACP method could then be applied to surface PDEs involving the surface gradient, Laplace-Beltrami operator, or surface divergence (see Appendix \ref{APP:numerical_details}) on arbitrary shapes.  In particular, the method could be applied to vesicles that are not axisymmetric, such as those under flow or in an arbitrary electric field.

\section*{Acknowledgements}
  This work was partially performed under the auspices of the U.S. Department of Energy by Lawrence Livermore National Laboratory under contract DE-AC52-07NA27344, Lawrence Livermore National Security, LLC.	The author acknowledges the valuable input from the reviewers that led to an improvement in the implementation of the CACP method, as well as funding support from the National Science Foundation [DMS-1304081].

%%%%%%%%%%%%%%%%%%%%%
%%%%%%%%%%%%%%%%%%%%%
\newpage
\appendix

\section{Lack of Side Condition for the CACP Method}
\label{APP:side_condition}

Asymptotic expansions are used to illustrate why the CACP method does not require a side condition like the CP method.  Recall the alternative PDE in $\mathbb{R}^2$ for the CACP method:
\begin{gather*}
	\big(1 + \phi(\mbfx)\kappa(\mbfx)\big) \nabla \cdot \big((1 + \phi(\mbfx)\kappa(\mbfx)) \nabla \gamma(\mbfx)\big) - c(\mbfx)\gamma(\mbfx) = f(\mbfx), \quad \mbfx \in \Omega_\Gamma
\end{gather*}
Given the surface definition $\Gamma = \{ \mbfr(\sigma) : \sigma \in [0,\Sigma) \}$, for some $\Sigma \in \mathbb{R}$, use the coordinate system $\mbfx(\sigma,\eta) = \mbfr(\sigma) + \eta \mbfn(\sigma)$ to define $\Omega^*_\Gamma = \{\mbfx(\sigma,\eta) : \sigma \in [0,\Sigma), \eta \in (\text{-}\delta,\delta)\}$ for some $\delta$ that ensures $\Omega^*_\Gamma \subset \Omega_\Gamma$.  Assume that $c(\mbfx)=c$ and $\nabla \gamma \cdot \mbfn = 0$ for $\mbfx \in \Omega_\Gamma \backslash \Omega^*_\Gamma$ so that the PDE to solve is
\begin{gather*}
	|\mbfr_\sigma(\sigma)|^{\text{-}1}\big(|\mbfr_\sigma(\sigma)|^{\text{-}1} \gamma_\sigma(\sigma,\eta)\big)_\sigma + \big((1 + \eta \kappa(\sigma))^2 \gamma_\eta(\sigma,\eta) \big)_\eta - c\gamma(\sigma,\eta) = f(\sigma), \,\, (\sigma,\eta) \in [0,\Sigma] \times (\text{-}\delta,\delta)\\
	\gamma_\eta(\sigma, \pm\delta) = 0
\end{gather*}
Numerically speaking, $\delta$ is chosen so $\Omega^*_\Gamma$ includes all grid points required to interpolate for $\gamma$ along the interface.  Thus, as the grid becomes more refined, $\delta \rightarrow 0$, and $\eta = \epsilon \tilde{\eta}$ for some small parameter $\delta$ (relabeled $\epsilon$) and $\mathcal{O}(1)$ quantity $\tilde{\eta}$.  Substituting in, the PDE now has a small parameter to expand about:
\begin{gather*}
	|\mbfr_\sigma|^{\text{-}1}\big(|\mbfr_\sigma|^{\text{-}1} \gamma_\sigma\big)_\sigma + \frac{1}{\epsilon^2}\big((1 + \epsilon \tilde{\eta} \kappa)^2 \gamma_{\tilde{\eta}} \big)_{\tilde{\eta}} - c\gamma = f, \quad (\sigma,\tilde{\eta}) \in [0,\Sigma) \times (\text{-}1,1)\\
	\gamma_{\tilde{\eta}}(\sigma, \pm 1) = 0
\end{gather*}
\par
Assume the expansion $\gamma(\sigma,\tilde{\eta}) \sim \gamma_0(\sigma,\tilde{\eta}) + \epsilon \gamma_1(\sigma,\tilde{\eta}) + \ldots + \epsilon^N \gamma_N(\sigma,\tilde{\eta})$, for some integer $N > 0$.  Substituting this into the above PDE leads to the following $\mathcal{O}(1)$ problem and solution:
\begin{gather*}
	\left .
	\begin{gathered}
		(\gamma_0)_{\tilde{\eta}\tilde{\eta}}(\sigma,\tilde{\eta}) = 0, \quad (\sigma,\tilde{\eta}) \in [0,\Sigma) \times (\text{-}1,1)\\
		(\gamma_0)_{\tilde{\eta}}(\sigma, \pm 1) = 0
	\end{gathered}
	\right \} \Rightarrow
	\begin{gathered}
		\gamma_0(\sigma,\tilde{\eta}) = b_0(\sigma).
	\end{gathered}
\end{gather*}
The $\mathcal{O}(\epsilon)$ problem and solution are identical to the $\mathcal{O}(1)$ problem, giving $\gamma_1(\sigma,\tilde{\eta}) = b_1(\sigma)$.  The coupling between expansion terms comes in the $\mathcal{O}(\epsilon^2)$ problem:
\begin{gather*}
	|\mbfr_\sigma(\sigma)|^{\text{-}1}\big(|\mbfr_\sigma(\sigma)|^{\text{-}1} (b_0)_\sigma(\sigma)\big)_\sigma + (\gamma_2)_{\tilde{\eta}\tilde{\eta}}(\sigma,\tilde{\eta}) - cb_0(\sigma) = f(\sigma), \quad (\sigma,\tilde{\eta}) \in [0,\Sigma) \times (\text{-}1,1)\\
	(\gamma_2)_{\tilde{\eta}}(\sigma, \pm 1) = 0
\end{gather*}
The general solution is $\gamma_2 =(1/2)\big[f + cb_0 - |\mbfr_\sigma|^{\text{-}1}\big(|\mbfr_\sigma|^{\text{-}1}(b_0)_\sigma\big)_\sigma \big]\tilde{\eta}^2 + a_2(\sigma) \tilde{\eta} + b_2(\sigma)$.  Enforcing the boundary condition determines $b_0(\sigma)$ and the particular solution for $\gamma_2(\sigma,\tilde{\eta})$:
\begin{gather*}
	\begin{gathered}
		|\mbfr_\sigma|^{\text{-}1}\big(|\mbfr_\sigma|^{\text{-}1}(b_0)_\sigma\big)_\sigma - cb_0 = f,\\
		b_0(0) = b_0(\Sigma), \quad\quad (b_0)_\sigma(0) = (b_0)_\sigma(\Sigma),
	\end{gathered}
	\quad\quad \text{and} \quad\quad \gamma_2(\sigma,\tilde{\eta}) = b_2(\sigma).
\end{gather*}
The last problem that needs examination before a generalization is made is the $\mathcal{O}(\epsilon^3)$ problem:
\begin{gather*}
	|\mbfr_\sigma(\sigma)|^{\text{-}1}\big(|\mbfr_\sigma(\sigma)|^{\text{-}1} (b_1)_\sigma(\sigma)\big)_\sigma + (\gamma_3)_{\tilde{\eta}\tilde{\eta}}(\sigma,\tilde{\eta}) - cb_1(\sigma) = 0, \quad (\sigma,\tilde{\eta}) \in [0,\Sigma] \times (\text{-}1,1)\\
	(\gamma_3)_{\tilde{\eta}}(\sigma, \pm 1) = 0
\end{gather*}
Enforcing the boundary condition on the associated general solution for $\gamma_3$ gives the following: $b_1(\sigma) = \alpha_1 \cosh\big(\sqrt{c}\,s(\sigma)\big) + \beta_1 \sinh \big(\sqrt{c}\,s(\sigma) \big)$, where $s(\sigma) = \int_0^\sigma |\mbfr_{\tilde{\sigma}}(\tilde{\sigma})| d\tilde{\sigma}$.  Enforcing the periodic boundary conditions on $b_1$ determines $b_1(\sigma) = 0$ and $\gamma_3(\sigma,\eta) = b_3(\sigma)$.
\par
The $\mathcal{O}(\epsilon^4), \mathcal{O}(\epsilon^5), \ldots, \mathcal{O}(\epsilon^N)$ problems follow that of the $\mathcal{O}(\epsilon^3)$ problem.  Thus, the asymptotic solution is
\begin{gather}
	\gamma(\sigma,\tilde{\eta}) \sim b_0(\sigma) + \epsilon^{N-1} b_{N-1}(\sigma) + \epsilon^N b_N(\sigma), \text{ where } \Delta_s b_0 - cb_0 = f, \label{EQ:asymptotic}
\end{gather}
for arbitrarily large $N$.  The same approach will not result in something similar to (\ref{EQ:asymptotic}) for the CP method.  The corresponding PDE with a small parameter for the CP method is
\begin{gather*}
	\frac{|\mbfr_\sigma|^{\text{-}1}}{1 + \epsilon \tilde{\eta}\kappa}\bigg(\frac{|\mbfr_\sigma|^{\text{-}1}}{1 + \epsilon \tilde{\eta} \kappa} \gamma_\sigma \bigg)_\sigma + \frac{1}{1 + \epsilon \tilde{\eta}\kappa} \bigg((1 + \epsilon \tilde{\eta} \kappa) \gamma_{\tilde{\eta}}\bigg)_{\tilde{\eta}} - c \gamma = f, \quad (\sigma,\tilde{\eta}) \in [0,\Sigma) \times (\text{-}1,1),\\
	\gamma_{\tilde{\eta}}(\sigma, \pm 1) = 0.
\end{gather*}
Assuming the same expansion as before, one finds that $\gamma_0(\sigma,\tilde{\eta}) = \gamma_0(\sigma)$, with $\gamma_0$ entirely determined by the $\mathcal{O}(1)$ problem.  One then looks at the $\mathcal{O}(\epsilon)$ problem:
\begin{gather*}
	|\mbfr_\sigma|^{\text{-}1}\bigg(|\mbfr_\sigma|^{\text{-}1} (\gamma_1)_\sigma \bigg)_\sigma +
	 (\gamma_1)_{\tilde{\eta}\tilde{\eta}} - c \gamma_1 = \tilde{\eta}\bigg( |\mbfr_\sigma|^{\text{-}1}\big(|\mbfr_\sigma|^{\text{-}1} \kappa (\gamma_0)_\sigma \big)_\sigma + \kappa \big(f + c \gamma_0\big)\bigg).
\end{gather*}
In order for $\gamma_1$ to be independent of $\tilde{\eta}$, the right-hand side must be zero.  Unless $f$ is specifically chosen to make this true for a particular $\kappa$, or vice versa, the term $\gamma_1$ must depend on $\tilde{\eta}$.

%%%%%%%%%%%%%%%%%%%%%
\section{Condition Numbers for the Comparison Examples}
\label{APP:condition_numbers}

\begin{table}[h]
	\center
	\begin{tabular}{|l||
		S[round-mode=places,round-precision=4]|
		S[round-mode=places,round-precision=4]|}
		\hline
		& {CP method} & {CACP method} \\
		\hline
		{$M = 80$} & 2.5643243749762e4 & 1.8254602642126e4 \\
		{$M = 160$} & 1.01872387028823e5 & 0.75342598827694e5 \\
		{$M = 320$} & 4.09613369447210e5 & 3.10202797309217e5 \\
		{$M = 640$} & 1.646390801857933e6 & 1.250179726914625e6 \\
		\hline
	\end{tabular}
	\caption{condition number estimates for the CP and CACP methods on the unit circle in $\mathbb{R}^2$}
	\label{TAB:condition_numbers_circle}
\end{table}

\begin{table}[h]
	\center
	\begin{tabular}{|l||
		S[round-mode=places,round-precision=4]|
		S[round-mode=places,round-precision=4]|}
		\hline
		& {CP method} & {CACP method} \\
		\hline
		{$M = 80$} & 2.7552687828114e4 & 2.4622776065804e4 \\
		{$M = 160$} & 1.11806458152813e5 & 0.88592266379691e5 \\
		{$M = 320$} & 4.23433628575171e5 & 3.14586335519011e5 \\
		{$M = 640$} & 1.646567925606908e6 & 1.186770882829378e6 \\
		\hline
	\end{tabular}
	\caption{condition number estimates for the CP and CACP methods on the clover in $\mathbb{R}^2$}
	\label{TAB:condition_numbers_clover}
\end{table}

\begin{table}[h]
	\center
	\begin{tabular}{|l||
		S[round-mode=places,round-precision=4]|
		S[round-mode=places,round-precision=4]|}
		\hline
		& {CP method} & {CACP method} \\
		\hline
		{$M = 40$} & 1.8703980422981e4 & 0.95450358517746e4 \\
		{$M = 80$} & 7.1973688729913e4 & 3.42552441831559e4 \\
		{$M = 160$} & 2.88559823145381e5 & 1.404234973213166e5 \\
		{$M = 320$} & 1.155944626032292e6 & 0.5849543773817249e6 \\
		\hline
	\end{tabular}
	\caption{condition number estimates for the CP and CACP methods on the sphere in $\mathbb{R}^3$}
	\label{TAB:condition_numbers_sphere}
\end{table}

\section{Inextensibility Condition for Axisymmetric Surfaces} \label{APP:inextensibility}

The inextensibility of the vesicle membrane is captured in the condition $\mbfr_{\sigma_i} g^{ij} \tilde{\mbfu}_{\sigma_j}$ from Seifert \cite{seifert_fluid_1999}.  Given an axisymmetric surface and velocity, this condition simplifies to (\ref{EQ:inextensibility}).  Let $\mbfr(\sigma,\theta) = x(\sigma)\mathbf{e}_x(\theta) + y(\sigma) \mathbf{e}_y$ describe the surface $\Gamma$, where $\mathbf{e}_x(\theta) = \cos(\theta) \mathbf{i} + \sin(\theta) \mathbf{j}$, $\mathbf{e}_y = \mathbf{k}$, and $\theta \in [0,2\pi)$.  Denote $\mbft^\sigma = \mbfr_\sigma / |\mbfr_\sigma|$ and $\mbft^\theta = \mbfr_\theta / |\mbfr_\theta|$.  Noting that $\tilde{\mbfu} \cdot \mbft^\theta = 0$ from the axisymmetry, the inextensibility condition can be expressed as
\begin{align*}
	&\frac{1}{|\mbfr_\sigma|} \tilde{\mbfu}_\sigma \cdot \mbft^\sigma + \frac{1}{|\mbfr_\theta|}\tilde{\mbfu}_\theta \cdot \mbft^{\theta} = 0\\
	\Rightarrow &\frac{1}{|\mbfr_\sigma|}\big((\tilde{\mbfu} \cdot \mbft^\sigma) \mbft^\sigma + (\tilde{\mbfu} \cdot \mbfn) \mbfn \big)_\sigma \cdot \mbft^\sigma + \frac{1}{|\mbfr_\theta|}\big((\tilde{\mbfu} \cdot \mbft^\sigma) \mbft^\sigma + (\tilde{\mbfu} \cdot \mbfn) \mbfn\big)_\theta \cdot \mbft^{\theta} = 0\\
\Rightarrow &\frac{1}{|\mbfr_\sigma|}(\tilde{\mbfu} \cdot \mbft^\sigma)_\sigma + (\tilde{\mbfu} \cdot \mbfn)\kappa + \frac{1}{|\mbfr_\theta|}(\tilde{\mbfu} \cdot \mbft^\sigma) \mbft^\sigma_\theta \cdot \mbft^\theta + (\tilde{\mbfu} \cdot \mbfn)h = 0\\
\Rightarrow &\frac{1}{|\mbfr_\sigma|}(\tilde{\mbfu} \cdot \mbft^\sigma)_\sigma + \frac{x_\sigma(\sigma)}{|\mbfr_\sigma| x(\sigma)}(\tilde{\mbfu} \cdot \mbft^\sigma) + H(\tilde{\mbfu} \cdot \mbfn) = 0\\
\Rightarrow &\frac{1}{x(\sigma)}\frac{1}{|\mbfr_\sigma|}\big(x(\sigma)(\tilde{\mbfu} \cdot \mbft^\sigma)\big)_\sigma + H(\tilde{\mbfu} \cdot \mbfn) = 0
\end{align*}

%%%%%%%%%%%%%%%%%%%%%
\section{Numerical Details for the Relaxing Vesicle Results} \label{APP:numerical_details}

The computational domain for the relaxing vesicle is discretized into grid nodes and grid cells.  Each grid node $\mbfx_{i,j}$ is a corner for four square grid cells, whose centers are $\mbfx_{i\pm1/2,j\pm1/2}$.  Note this means there is a layer of ghost grid cells around the entire computational domain.  Additionally, a layer of ghost grid nodes is added at $x=0$ to enforce a computational boundary condition of axisymmetry there.  Most quantities reside at the grid nodes, except for the stream functions and pressures that reside at the grid cell centers.  The finite-difference stencils across the grid are mostly taken from Vogl \cite{vogl_curvature-augmented_2016}, with adjustments needed to account for singularities that result from the cylindrical coordinate system.
\par
In computing $\psi_f$ from (\ref{EQ:decomp_f}), the advection term $\nabla \mbfu . \mbfu$ has the same form as when using two-dimensional Cartesian coordinates and uses the same ENO scheme as in \cite{vogl_curvature-augmented_2016}.  The vector Laplacian has a different form, however, and requires special treatment.  Denote $\mbfu(\mbfx,t) = u(\mbfx,t) \mathbf{e}_x + v(\mbfx,t) \mathbf{e}_y$ so that the vector Laplacian is approximated as
\begin{align*}
	\Delta \mbfu(\mbfx_{i,j},t) =& \bigg[ \left(\frac{1}{x} (x u)_x \right)_x + u_{yy} \bigg] \mathbf{e}_x + \bigg[ \frac{1}{x_{i,j}} (x v_x)_x + v_{yy}\bigg] \mathbf{e}_y \\
	\approx& \bigg[ \frac{(xu)_{i+1,j} - (xu)_{i,j}}{x_{i+1/2,j}\Delta x^2}  - \frac{(xu)_{i,j} - (xu)_{i-1,j}}{x_{i-1/2,j}\Delta x^2} +   \frac{u_{i,j+1} - 2u_{i,j}+ u_{i,j-1} }{\Delta x ^2} \bigg] \mathbf{e}_x \\
	&+ \bigg[ \frac{x_{i,j+1/2}(v_{i,j+1} - v_{i,j}) - x_{i,j-1/2}(v_{i,j} - v_{i,j-1})}{x_{i,j}\Delta x^2} + \frac{v_{i,j+1} - 2v_{i,j}+ v_{i,j-1} }{\Delta x ^2} \bigg] \mathbf{e}_y.
\end{align*}
If $x_{i,j} = 0$, then it is assumed that $v(x,y,t) \approx v_{i,j} + (v_{i+1,j} - v_{i,j}) x^2/\Delta x^2$ in a neighborhood of $x_{i,j}$.  This allows for $(x v_x)_x / x$ to be approximated by $4 (v_{i+1,j} - v_{i,j} ) / \Delta x^2$:
\begin{align*}
	\Delta \mbfu(\mbfx_{i,j},t) \approx& \bigg[ \frac{(xu)_{i+1,j} - (xu)_{i,j}}{x_{i+1/2,j}\Delta x^2}  - \frac{(xu)_{i,j} - (xu)_{i-1,j}}{x_{i-1/2,j}\Delta x^2} +   \frac{u_{i,j+1} - 2u_{i,j}+ u_{i,j-1} }{\Delta x ^2} \bigg] \mathbf{e}_x \\
	&+ \bigg[ 4 \frac{ v_{i+1,j} - v_{i,j} }{\Delta x^2} + \frac{v_{i,j+1} - 2v_{i,j}+ v_{i,j-1} }{\Delta x ^2} \bigg] \mathbf{e}_y.
\end{align*}
Recall that $\mathbf{e}_\theta(\theta) = -\sin(\theta) \mathbf{e}_x(\theta) + \cos(\theta) \mathbf{e}_y$ and that the Poisson equation for $\psi_f$ comes from a dot product taken after a cross product of (\ref{EQ:decomp_f}).  Thus, the stencil for $\psi_f$ is
\begin{align*}
	\nabla \times &(\nabla \times (\psi_f \mathbf{e}_\theta) ) \cdot \mathbf{e}_\theta|_{\mbfx_{i,j}} = -\bigg[ \frac{1}{x}(x \psi_f)_x \bigg]_x|_{\mbfx_{i,j}} - (\psi_f)_{yy}|_{\mbfx_{i,j}}\\
	\approx& -\bigg[ \frac{(x\psi_f)_{i+1,j} - (x\psi_f)_{i,j}}{x_{i+1/2,j}\Delta x^2}  - \frac{(x\psi_f)_{i,j} - (x\psi_f)_{i-1,j}}{x_{i-1/2,j}\Delta x^2} +   \frac{(\psi_f)_{i,j+1} - 2(\psi_f)_{i,j}+ (\psi_f)_{i,j-1} }{\Delta x ^2} \bigg]\\
\end{align*}
If $x_{i-1/2,j} = 0$, then $\psi_f(x,y,t)$ is approximated in a neighborhood of $x_{i-1/2,j} = 0$ so that $[(x \psi_f)_x/x]_x$ can be evaluated:
\begin{align*}
	&\psi_f(x,y,t) \approx (\psi_f)_{i,j} \frac{x}{0.5 \Delta x} + \big((\psi_f)_{i+1,j} - 3(\psi_f)_{i,j}\big)\frac{x(x - 0.5 \Delta x)}{1.5\Delta x^2} \\
	\Rightarrow &\nabla \times (\nabla \times (\psi_f \mathbf{e}_\theta) ) \cdot \mathbf{e}_\theta|_{\mbfx_{i,j}} \approx -\bigg[ \frac{13/6(\psi_f)_{i+1,j} - 6.5(\psi_f)_{i,j}}{\Delta x^2} + \frac{(\psi_f)_{i,j+1} - 2(\psi_f)_{i,j}+ (\psi_f)_{i,j-1} }{\Delta x ^2} \bigg]
\end{align*}
The right-hand side of the Possion equation is handled as in \cite{vogl_curvature-augmented_2016}.
\par
Computing $\psi_b$ from (\ref{EQ:decomp_b}) and $\psi_\gamma$ from (\ref{EQ:decomp_g}) uses the same stencil for $\psi_b$ and $\psi_\gamma$ as above.  For the other terms, some additional work is needed.  The quantity $\Delta_s H$ is needed at the grid nodes, but cannot be evaluated directly from $H$ due to smoothness limitation of the fast marching method used in reinitialization (see \cite{vogl_curvature-augmented_2016}).   Instead, it is obtained from a surface divergence of $\nabla_s H$.  From inspection of the surface gradient and Laplace-Beltrami operators and their embeddings (\ref{EQ:LB_relation3D_axis}), the form of the surface divergence and its generalized embedding are
\begin{align*}
	\nabla_s \cdot \mathbf{f} := \frac{1}{x(\sigma)}\frac{1}{|\mbfr_\sigma|} \big( x(\sigma) \mathbf{f} \cdot \mbft^\sigma\big)_\sigma  = (1 + \phi \kappa)(1 + \phi h)\nabla \cdot \left(\frac{\mathbf{f}}{1 + \phi h} \right).
\end{align*}
Denote $\mbfn(\mbfx,t) = n^x(\mbfx,t)\mathbf{e}_x + n^y(\mbfx,t)\mathbf{e}_y$ and $\mbft = -n^y \mathbf{e}_x + n^x \mathbf{e}_y$.  The finite-difference stencil for evaluating $\Delta_s H$ from $H_s$ values, which are available from the reintialization process, is
\begin{align*}
	\nabla \cdot \left(\frac{H_s \mbft}{1 + \phi h} \right)(\mbfx_{i,j}) \approx& \frac{1}{x_{i,j}\Delta x}\bigg [ x_{i+1,j}\left(\frac{-H_s n^y}{1 + \phi h}\right)_{i+1,j}  - x_{i-1,j}\left(\frac{-xH_s n^y}{1 + \phi h}\right)_{i-1,j}\bigg] \\
	&+ \frac{1}{\Delta x}\bigg[ \left(\frac{H_s n^x}{1 + \phi h}\right)_{i,j+1}  - \left(\frac{H_s n^x}{1 + \phi h}\right)_{i,j-1}\bigg].
\end{align*}
If $x_{i,j}=0$, then $[$-$H_s n^y/(1 + \phi h)](x,y,t)$ is approximated in a neighborhood of $x_{i,j} = 0$ so that the corresponding terms can be evaluated numerically:
\begin{align*}
	&\left(\frac{-H_s n^y}{1 + \phi h}\right)(x,y,t) \approx \left( \frac{-H_s n^y}{1 + \phi h}\right)_{i+1,j}\frac{x}{\Delta x}\\
	\Rightarrow & \nabla \cdot \left(\frac{H_s \mbft}{1 + \phi h} \right)(\mbfx_{i,j},t) \approx \frac{2}{\Delta x}\left(\frac{-H_s n^y}{1 + \phi h}\right)_{i+1,j} + \frac{1}{\Delta x}\bigg[ \left(\frac{H_s n^x}{1 + \phi h}\right)_{i,j+1}  - \left(\frac{H_s n^x}{1 + \phi h}\right)_{i,j-1}\bigg].
\end{align*}
Finally, denote $B = \text{Be}(\Delta_s H - 2KH + 0.5H^3)$.  This gives the following for the right-hand side of the Poisson equation for $\psi_b$:
\begin{align*}
	\nabla \times \big( \delta(\phi)B\mbfn \big) \cdot \mathbf{e}_\theta &= \big[ \nabla \big( \delta(\phi) \big) \times B\mbfn + \delta(\phi) \nabla B \times \mbfn + \delta(\phi) B \nabla \times \mbfn \big] \cdot \mathbf{e}_\theta \\
&= \delta(\phi) \big(-B_x n^y + B_y n^x \big) = \delta(\phi) \nabla B \cdot \mbft.
\end{align*}
For the right-hand side of the Poisson equation for $\psi_\gamma$,
\begin{align*}
	\nabla \times \big( \delta(\phi) (\gamma_s \mbft - \gamma H \mbfn) \big) \cdot \mathbf{e}_\theta =& \big[ \nabla \big( \delta(\phi) \big) \times (\gamma_s \mbft - \gamma H \mbfn) + \delta(\phi)\big( \nabla (\gamma_s) \times \mbft \\
&+ \gamma_s \nabla \times \mbft - \nabla (\gamma H) \times \mbfn - \gamma H \nabla \times \mbfn \big) \big] \cdot \mathbf{e}_\theta \\
=& \left(\delta'(\phi) + \delta(\phi) \frac{\kappa}{1 + \phi \kappa} \right) \gamma_s + \delta(\phi) \nabla (\gamma H) \cdot \mbft,
\end{align*}
which used that $\nabla \times \mbft = \kappa / (1 + \phi \kappa)$.  The gradient of $B$ and of $\gamma H$ are discretized as in \cite{vogl_curvature-augmented_2016}.
\par
The next step in the numerical approach that needs elaboration is the simultaneous solving of $\gamma$ and $p_\gamma$.  Recall that the Poisson equation for $p_\gamma$ comes from a divergence of (\ref{EQ:decomp_g}).  The stencil for $\Delta p_\gamma$ is the same as for the $\mathbf{e}_y$ component of $\Delta \mbfu$.  Note that no singularity is involved, because $p_\gamma$ resides at the cell centers.  The remaining terms are then
\begin{align*}
	\nabla \cdot \big[ \delta(\phi)\big(\gamma_s \mbft - \gamma H \mbfn\big)\big] &= -\delta'(\phi) \gamma H + \delta(\phi) \big[\nabla \cdot (\gamma_s \mbft) - \gamma H \nabla \cdot \mbfn \big] \\
&= -\big[ \delta'(\phi) + \nabla \cdot \mbfn \big] H \gamma + \delta(\phi) \nabla \cdot \big( (1 + \phi\kappa) \nabla \gamma \big), \text{ where} \\
\nabla \cdot \mbfn = \frac{1}{x}n^x + n^x_x + n^y_y &= \frac{1}{x}\frac{y_\sigma(\sigma)}{|\mbfr_\sigma|} + \frac{\kappa}{1 + \phi \kappa} = \frac{x(\sigma)}{x}h + \frac{\kappa}{1 + \phi\kappa} = \frac{x - \phi n^x}{x}h + \frac{\kappa}{1 + \phi \kappa}.
\end{align*}
Note that because $p_\gamma$ resides on cell centers, the grid nodes quantities $\big(\delta'(\phi) + \nabla \cdot \mbfn \big) H$ and $\delta(\phi)$ are interpolated using a bi-linear stencil.  The variable-coefficient Laplacian stencil for the remaining term is
\begin{align*}
	\nabla \cdot \big( (1 + \phi \kappa) \nabla \gamma)(\mbfx_{i,j},t) \approx& \frac{1}{x_{i,j}}\frac{(1 + \phi \kappa)_{i+1,j} x_{i+1,j} (\gamma_x)_{i+1,j} - (1 + \phi \kappa)_{i-1,j} x_{i-1,j}(\gamma_x)_{i-1,j}}{2\Delta x} \\
&+ \frac{(1 + \phi \kappa)_{i,j+1} (\gamma_y)_{i,j+1} - (1 + \phi \kappa)_{i,j-1} (\gamma_y)_{i,j-1}}{ 2\Delta x},
\end{align*}
where the grid quantity $1 + \phi \kappa$ is also evaluated at cell centers $\mbfx_{i \pm1, j \pm1}$ with bi-linear interpolation.  The terms $(\gamma_x)$ and $(\gamma_y)$ are discretized as regular Cartesian derivatives \cite{vogl_curvature-augmented_2016}.
\par
Recall that surface PDE to solve for $\gamma$ comes from enforcing (\ref{EQ:inextensibility}) on $\tilde{\mbfu}^{n+1/2}$ after denoting $\mbfu^* = \mbfu^n + 0.5(\Delta t / \text{Re})[\nabla \times \Psi_f^n + \nabla \times \Psi_b^{n+1/2}]$.  This comes from evaluating the velocity update equation at $\mbfr(\sigma,t_{n+1/2})$:
\begin{align*}
	\tilde{\mbfu}^{n+1/2} &= \mbfu^*(\mbfr(\sigma,t_{n+1/2})) + \frac{\Delta t}{2\text{Re}}\big[ \delta(0)(\gamma_s \mbft^\sigma - H \gamma \mbfn)^{n+1/2}(\sigma) - \nabla p_\gamma^{n+1/2}(\mbfr(\sigma,t_{n+1/2}))\big] \\
	\Rightarrow &0 = \frac{1}{x(\sigma)}\frac{1}{|\mbfr_\sigma|}\big[ x(\sigma) \big(\mbfu^*(\mbfr(\sigma,t_{n+1/2})) \cdot 		\mbft^\sigma \big) \big]_\sigma + H^{n+1/2} \big[ \mbfu^*(\mbfr(\sigma,t_{n+1/2})) \cdot \mbfn^{n+1/2} \big] \\
	&+ \frac{\Delta t}{2 \text{Re}} \big[ \delta(0) (\Delta_s \gamma - H^2 \gamma)^{n+1/2}(\sigma) - \frac{1}{x(\sigma)}\frac{1}{|\mbfr_\sigma|} \big[ x(\sigma) \big( \nabla p_\gamma^{n+1/2}(\mbfr(\sigma,t_{n+1/2})) \cdot \mbft^\sigma \big)\big] \\
	&- H^{n+1/2} \big[ \nabla p_\gamma(\mbfr(\sigma,t_{n+1/2})) \cdot \mbfn^{n+1/2} \big] \\
	\Rightarrow & H^2 \gamma - \Delta_s \gamma + \frac{1}{\delta(0)}\left [\frac{1}{x}\big( x (\tilde{\nabla p}_\gamma \cdot \mbft^\sigma ) \big)_s + H (\tilde{\nabla p}_\gamma \cdot \mbfn)\right] \\
	&= \frac{2\text{Re}}{\delta(0)\Delta t} \bigg[ \frac{1}{x}\big( x (\tilde{\mbfu}^* \cdot \mbft^\sigma) \big)_s + H(\tilde{\mbfu}^* \cdot \mbfn) \bigg],
\end{align*}
where the $n+1/2$ notation is only dropped for notational brevity.  The Laplace-Beltrami term $\Delta_s \gamma$ is replaced by the proper embedding (\ref{EQ:LB_relation3D_axis}) and discretized using the finite-difference stencil
\begin{align*}
	\nabla \cdot \left( \frac{1 + \phi \kappa}{1 + \phi h} \nabla \gamma \right)(\mbfx_{i,j},t) \approx& \frac{1}{x_{i,j}} \bigg[ \bigg(x\frac{1 + \phi \kappa}{1 + \phi h} \bigg)_{x+} \frac{\gamma_{i+1,j} - \gamma_{i,j}}{\Delta x^2} - \bigg(x\frac{1 + \phi \kappa}{1 + \phi h} \bigg)_{x-} \frac{\gamma_{i,j} - \gamma_{i-1,j}}{\Delta x^2}\bigg] \\
&+ \bigg[ \bigg(\frac{1 + \phi \kappa}{1 + \phi h} \bigg)_{y+} \frac{\gamma_{i,j+1} - \gamma_{i,j}}{\Delta x^2} - \bigg(\frac{1 + \phi \kappa}{1 + \phi h} \bigg)_{y-} \frac{\gamma_{i,j} - \gamma_{i,j-1}}{\Delta x^2}\bigg],
\end{align*}
where a subscript of $x\pm$ denotes the average of that term's value at $x_{i\pm1,j}$ and $x_{i,j}$.  Similarly, a subscript of $y\pm$ denotes the average of that term's value at $y_{i,j\pm1}$ and $y_{i,j}$.  The singularity at $x_{i,j} = 0$ is handled in the same manner as the $\mathbf{e}_y$ term in $\Delta \mbfu$.  For the pressure, note the associated differential operator is the Laplace-Beltrami operator, so that
\begin{align*}
	\frac{1}{x}\big( x (\tilde{\nabla p}_\gamma \cdot \mbft^\sigma ) \big)_s = \frac{1}{x}\bigg( x \big[p_\gamma(\mbfr(\sigma,t),t)\big]_s \bigg)_s =  (1 + \phi \kappa)(1 + \phi h) \nabla \cdot \left( \frac{1 + \phi \kappa}{1 + \phi h} \nabla \big(p_\gamma |_{\cp(\mbfx) }\big) \right).
\end{align*}
Thus, the stencil for the pressure is the stencil for $\Delta_s \gamma$ combined with bi-cubic interpolation to evaluate the terms at $\cp(\mbfx_{k,l})$ instead of $\mbfx_{k,l}$.  Finally, the associated differential operator for $\tilde{\mbfu}$ is the surface divergence, so that
\begin{align*}
	\frac{1}{x}\big( x (\tilde{\mbfu}^* \cdot \mbft^\sigma) \big)_s = \frac{1}{x}\bigg( x \big[\mbfu(\mbfr(\sigma,t),t\big)^* \cdot \mbft^\sigma \big] \bigg)_s = (1 + \phi \kappa)(1 + \phi h)\nabla \cdot \left( \frac{ \mbfu |_{\cp(\mbfx)} }{1 + \phi h} \right).
\end{align*}
Thus, the stencil for $\Delta_s H = \nabla_s \cdot (\nabla_s H)$ is used after also being combined with a bi-cubic interpolation stencil.

\newpage
\bibliography{manuscript}
\bibliographystyle{abbrv}

\end{document}